\documentclass[10pt, journal, final, twocolumn, a4paper]{IEEEtran}

\usepackage{epsfig} 
\usepackage{graphicx}   
\usepackage{cite}
\usepackage{amssymb}
\usepackage{times}
\usepackage{subfigure}
\usepackage{placeins}
\usepackage{float}

\usepackage{amsmath}
\usepackage{tabularx}

\usepackage{setspace}
\usepackage{umoline}
\usepackage[dvips]{color}

\usepackage[outerbars,color]{changebar}
\cbcolor{red} 
\usepackage{relsize}
\usepackage{stfloats}

\begin{document}
\bibliographystyle{IEEEtran}

\title{Is the Low-Complexity Mobile-Relay-Aided FFR-DAS Capable of Outperforming the High-Complexity CoMP?}
\author{Shaoshi~Yang, Xinyi~Xu, Dimitrios~Alanis, Soon~Xin~Ng and~Lajos~Hanzo,~\IEEEmembership{Fellow,~IEEE}
\thanks{{The financial support of the Research Councils UK (RCUK) under the auspices of the India-UK Advanced Technology Centre (IU-ATC),  of the European Research Council's Advanced Fellow Grant, of the EU under the auspices of the CONCERTO project, and of the Chinese National 863 Programme project (No. 2015AA10302) is gratefully acknowledged.}

S. Yang, D. Alanis, S. X. Ng and L. Hanzo are with the School 
of Electronics and Computer Science, University of Southampton, 
Southampton, SO17 1BJ, UK (e-mail: \{sy7g09, da4g11, sxn, lh\}@ecs.soton.ac.uk).

X. Xu is with the China Academy of Electronics and Information Technology, P. R. China, and was with the School 
of Electronics and Computer Science, University of Southampton, 
Southampton, SO17 1BJ, UK (e-mail: xuxinyi718@sina.com).
}}

\markboth{The work has been accepted to publish on IEEE Transactions on Vehicular Technology, Mar. 2015.}%
{Shell \MakeLowercase{\textit{et al.}}: Bare Demo of IEEEtran.cls
for Journals}
\maketitle

\begin{abstract}
Coordinated multi-point transmission/reception aided collocated antenna system (CoMP-CAS) and mobile relay assisted fractional frequency reuse distributed antenna system (MR-FFR-DAS) constitute a pair of virtual-MIMO based technical options for achieving high spectral efficiency in interference-limited cellular networks. In practice both techniques have their respective pros and cons, which are studied in this paper by evaluating the achievable cell-edge performance on the uplink of multicell systems. We show that assuming the same antenna configuration in both networks, the maximum available cooperative spatial diversity inherent in the MR-FFR-DAS is lower than that of the CoMP-CAS. However, when the cell-edge MSs have a low transmission power, the lower-complexity MR-FFR-DAS relying on the simple single-cell processing may outperform the CoMP-CAS by using the proposed soft-combining based probabilistic data association (SC-PDA) receiver, despite the fact that the latter scheme is more complex and incurs a higher cooperation overhead. Furthermore, the benefits of the SC-PDA receiver may be enhanced by properly selecting the MRs' positions. Additionally, we show that the performance of the cell-edge MSs roaming near the angular direction halfway between two adjacent RAs (i.e. the "worst-case direction") of the MR-FFR-DAS may be more significantly improved than that of the cell-edge MSs of other directions by using multiuser power control, which also improves the fairness amongst cell-edge MSs. Our simulation results show that given a moderate MS transmit power, the proposed MR-FFR-DAS architecture employing the SC-PDA receiver is capable of achieving significantly better bit-error rate (BER) and effective throughput across the entire cell-edge area, including even the ”worst-case direction” and the cell-edge boundary, than the CoMP-CAS architecture.
\end{abstract}


\section{Introduction}
\label{secUL_intro}
\IEEEPARstart{F}{uture} mobile communication systems are expected to provide higher data rates and more homogeneous quality of service (QoS) across the entire network. In order to meet these demands, various technical options have been suggested, which include but are not limited to using more spectrum (e.g. millimeter wave), using more antennas (e.g. massive multiple-input multiple-output (MIMO) systems) and introducing dedicated relays as well as small cells to form heterogeneous networks\cite{Andrews_2014:5G}. Although these options are promising, most of them require installing additional equipment and a radical change of the network architecture and entities, which may be costly for the operators. From a pragmatic perspective, in the short- and medium-term it may be more promising to upgrade the existing cellular architecture using an evolutionary strategy. Hence in this paper, we aim for investigating the pros and cons of a pair of representative cellular architectures, namely the 
coordinated multi-point transmission/reception\cite{xx08r:LTEA2010, 3GPP_LTE:CoMP, Sawahashi_2010:CoMP, Gesbert_2010:multicell_MIMO, Yang:BSCoop, BScp:Marsch, Fettweis_2011:CoMP, Lee_2012:CoMP_challenges, Caire_2011:coop_cellular,Caire_2013:JSDM_CoMP_capacity} aided collocated antenna system (CoMP-CAS) and the fractional frequency reuse\cite{optical:LTE, Saquib_2013:FFR_LTE, Ganti_2011:FFR_journal_analytical} assisted distributed antenna system relying on mobile relays (MR-FFR-DAS), in the context of the multicell uplink. Both of them have the potential of providing a significant gain without incurring dramatic changes of the existing cellular systems, hence they are of great interest to both industry and academia.

\textbf{Benefits and challenges of CoMP-CAS}:  On the one hand, the conventional cellular architecture that relies on a CAS at each base station (BS) is still widely used, where the mobile stations (MSs) roaming in the cell-edge area typically suffer from a low throughput and low power efficiency. This is because low signal-to-interference-plus-noise ratio (SINR) may be experienced by the cell-edge MSs owing to the combined effects of inter-cell co-channel interference (CCI) and pathloss. As a remedy, CoMP  techniques\cite{Gesbert_2010:multicell_MIMO, Yang:BSCoop, BScp:Marsch, Caire_2011:coop_cellular,Caire_2013:JSDM_CoMP_capacity, xx08r:LTEA2010, 3GPP_LTE:CoMP,Sawahashi_2010:CoMP, Fettweis_2011:CoMP, Lee_2012:CoMP_challenges} have been advocated for the CAS based cellular architecture in the 3GPP Long Term Evolution Advanced (LTE-A) standard, where the inter-cell CCI can be mitigated or even beneficially exploited by cooperation amongst the different sectors or cells. 

However, in order to enhance the uplink performance of the cell-edge MSs, a CoMP-aided CAS requires these MSs to transmit at a rather high power\cite{Yang:BSCoop, DAS:coop, BSCP:Simeone}. Essentially, CoMP on the uplink relies on the joint decoding philosophy and it achieves a cooperative spatial diversity gain with the aid of the collaborative adjacent BSs equipped with CASs, which are quite far from the cell-edge MSs. Additionally, the CoMP techniques typically require exchanging a significant amount of data/channel information amongst the cooperative entities, which results in a potentially excessive overhead traffic on the backhaul. Furthermore, practical impairments, such as the asynchronous nature of inter-cell CCI\cite{Hongyuan_2008:asynchronous_BS_coop}, the backhaul capacity limitation, the channel estimation inaccuracy and the low channel-coherence time of the network-wide system\cite{Fettweis_2011:CoMP, Lee_2012:CoMP_challenges, Caire_2011:coop_cellular,Caire_2013:JSDM_CoMP_capacity, Caire_2010:rethinking_network_MIMO, Adve_2014:massive_MIMO_vs_network_MIMO}, may also significantly degrade the practically achievable benefits of the CoMP techniques. 
 
\textbf{Benefits and challenges of MR-FFR-DAS}: On the other hand, the FFR philosophy~\cite{Haipeng_2007:FFR, optical:ffr, Stolyar_2008:FFR, Ganti_2011:FFR_journal_analytical, Saquib_2013:FFR_LTE}, which confines the geographic scattering of the inter-cell CCI at the cost of a moderately reduced degree of frequency reuse, has also been suggested for the LTE initiative~\cite{optical:LTE}. Furthermore, the large pathloss experienced by the cell-edge MSs may be reduced by employing DASs, where the remote antennas (RAs) are positioned closer to the cell-edge MSs and connected to the BS using optical fibre~\cite{xinyi:DASDL, optical:Wake2004,  optical:AndrewsDAS}. Thus, the cell-edge MSs can transmit at a relatively low power. Additionally, it is possible to invoke the existing MSs as MRs~\cite{5196673, BSCP:Simeone}, which may provide additional benefits (reduced pathloss and increased spatial diversity gain) for cell-edge MSs. Hence, for a cellular system which cannot afford the more complex BS-cooperation aided CoMP, we suggest that it might be a practically attractive solution to amalgamate the benefits of FFR, DAS and MRs in each single cell for the sake of increasing the SINR at the RAs (on the uplink) or at the cell-edge MSs (on the downlink). 

The MR-FFR-DAS also faces particular challenges imposed by the system architecture. In the MR-FFR-DAS each cell-edge MS is served \textit{mainly} by a nearby RA, while the other RAs cannot provide the same level of support, since they are far away from the cell-edge MS considered. As a result, although the cell-edge MSs roaming close to the RAs do indeed benefit from a high SINR, there exist undesirable scenarios, where these cell-edge MSs suffer from increased intra-cell CCI. More specifically, when a cell-edge MS roaming near the angular direction halfway between two adjacent RAs, the SINR at the MS (on the downlink) or at its serving RA (on the uplink) may be substantially degraded,\footnote{This is because the cell-edge MSs of the same cell are operating in the same frequency band in the MR-FFR-DAS.} which we refer to as the ``worst-case direction'' problem~\cite{xinyi:DASDL}. 

\textbf{Motivations for the comparative study and related work:} As detailed above, both the CoMP-CAS and the MR-FFR-DAS have their particular pros and cons, despite sharing a similar virtual MIMO model. In general, the former scheme is more complex and yet, its practically achievable performance may be disappointing, as demonstrated in\cite{Hongyuan_2008:asynchronous_BS_coop, Fettweis_2011:CoMP, Lee_2012:CoMP_challenges, Caire_2011:coop_cellular,Caire_2013:JSDM_CoMP_capacity, Caire_2010:rethinking_network_MIMO,  Adve_2014:massive_MIMO_vs_network_MIMO}. Hence in this paper, we aim for characterizing the cell-edge performance of the lower-complexity MR-FFR-DAS in the context of the multicell uplink, which has not been disseminated in the open literature before. Additionally, we aim to provide further insights into the question whether the MR-FFR-DAS relying on single-cell processing constitutes a promising technical option in the scenario considered. Naturally, holistic cellular system design hinges on numerous technical aspects and target specifications, hence it is a challenge to make ``absolutely fair'' comparisons between two system-level designs and there is usually no definitive answer to the question of ``which design is better''. 

The existing BS cooperation aided CoMP reception techniques conceived for the multicell uplink typically rely on the philosophies of either egoistic ``interference cancellation''\cite{Mayer:Turbo_BS_cooperation_interference_cancellation, BSCP:Simeone, Khattak:distributed_max_log_MAP} or altruistic ``knowledge sharing and data fusion'' amongst BSs~\cite{Yang:BSCoop, BSCP:Aktas}, while both philosophies impose different backhaul traffic requirements. In our previous work\cite{xinyi:DASDL}, we have shown that in the downlink of the FFR-DAS, the intra-cell CCI imposed by the RAs may be mitigated by transmit preprocessing (TPP) dispensing with high-complexity multicell cooperation. As a beneficial result, the downlink throughput and coverage-quality in the cell-edge area of a multicell multiuser network may be significantly improved, where each BS simply plays the role of the central signal processing unit (CSPU). By contrast, in the FFR-DAS uplink dispensing with multicell cooperation, the intra-cell CCI may be mitigated by single-cell multiuser detection (MUD) techniques. This is because the intra-cell CCI of the uplink is essentially constituted by the multiuser interference (MUI) and the RAs are all connected to the BS, which is capable of carrying out centralized joint reception. However, it should be noted that due to the geometry of the DAS -- some RAs are close to the given cell-edge MSs and others are far -- the average receive SINRs recorded at 
different RAs for a particular cell-edge MS are significantly different, which may cause the so-called ``near-far problem''. 

\textbf{Novel contributions:} For the sake of characterizing the achievable cell-edge performance of both the CoMP-CAS and of the MR-FFR-DAS in the multicell uplink, we consider a set of four advanced MUD-based reception techniques which are more robust to the near-far problem than linear MUDs. More explicitly, three non-cooperative single-cell MUDs, namely the classic minimum mean-square error (MMSE) based optimal user ordering aided successive interference cancellation (MMSE-OSIC)~\cite{5370668}, the ``exhaustive brute-force search" based maximum-likelihood detection (ML)~\cite{5288446} and the probabilistic data association (PDA)~\cite{Yang:nominal_to_true_APP} scheme are investigated for the single-cell processing that relies on neither BS cooperation nor MRs. Furthermore, a low-backhaul-traffic ``knowledge sharing and data fusion" based soft-combining PDA (SC-PDA)~\cite{Yang:BSCoop} scheme is conceived for both the CoMP-CAS and the MR-FFR-DAS. The novel contributions of this paper are summarized as follows.
  
\begin{enumerate}  
\item We demonstrate that the maximum achievable cooperative spatial diversity inherent in the MR-FFR-DAS is lower than that of the CoMP-CAS, when each BS of both systems has the same number of antennas serving single-antenna MSs and only a single MR is invoked for each cell-edge MS. Despite this, when assuming that the cell-edge MSs have a low transmission power and are located in the close vicinity of RAs, even the FFR-DAS invoking no MRs may outperform the CoMP-CAS.

\item We show that as an effective remedy to the above-mentioned ``worst-case direction'' problem, regardless of the MS positions the SC-PDA based receiver that relies on MRs is capable of providing a significant cooperative diversity gain compared to the non-cooperative MMSE-OSIC and PDA based MUDs that invoke no MRs. Furthermore, the benefits of the SC-PDA based receiver may be enhanced by carefully selecting the MRs' positions from an identified ``reliable area''.  

\item The QoS distribution of cell-edge MSs is visualized for the MR-FFR-DAS, which demonstrates that although power control is an inefficient technique in interference-limited scenarios, it is more useful for improving the cell-edge MSs' performance in the worst-direction than in the best-direction of the MR-FFR-DAS considered. This insight is valuable for improving the fairness amongst cell-edge MSs.    
\end{enumerate}

The rest of this paper is organized as follows. In Section~\ref{sec2UL}, we describe the multicell topology of both the CoMP-CAS and MR-FFR-DAS regimes. In Section~\ref{secUL_rs}, we detail the received signal models of both schemes. Then, in Section \ref{secUL_CP}, the set of four MUD-based cooperative/noncooperative reception schemes as well as the power control technique invoked are described. The performance comparison results of the CoMP-CAS and of the MR-FFR-DAS are presented in Section~\ref{secUL_PfEv}. Finally, our conclusions are offered in Section~\ref{secUL_cn}.

\section{System Description}
\label{sec2UL}

\subsection{Multicell Multiuser System Topology}
\subsubsection{MR-FFR-DAS Architecture}
\begin{figure}[tbp]
    \centering
    {
        \includegraphics[width=0.9\linewidth]{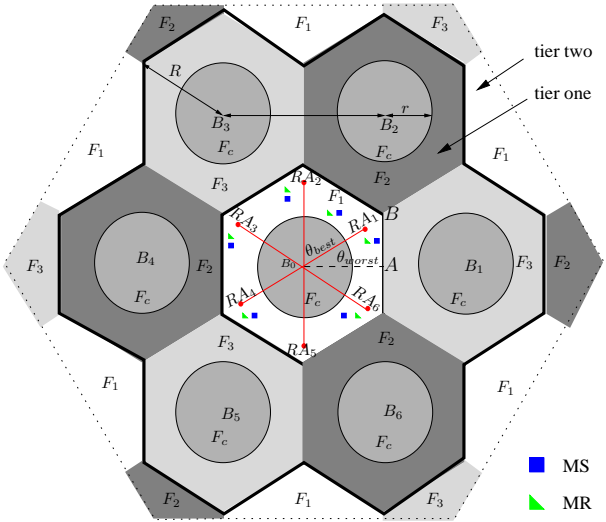}
    } 
    \caption{The cellular topology of the MR-FFR-DAS architecture, where $N_r = 6$ distributed antennas are employed and $N_t= 6$ MSs randomly roam in the cell-edge area.}
    \label{fig:Tpgy_DAS}
\end{figure}
The MR-FFR-DAS architecture supporting a multicell multiuser operating scenario~\cite{xinyi:DASDL} consists of two tiers of 19 hexagonal cells, as seen in Fig~\ref{fig:Tpgy_DAS}. The frequency partitioning strategy of the total available bandwidth $F$ is characterized by $F_{c}\cap F_{e}=\oslash$, where $F_c$ and $F_e$ represent the cell-centre's frequency band and the cell-edge's frequency band, respectively. Furthermore, $F_{e}$ is divided into three orthogonal frequency bands $F_i,i\in\{1,2,3\}$, which are exclusively used at the cell-edge of each of the three adjacent cells. We have demonstrated that this regime is capable of sufficiently reducing the inter-cell CCI in the FFR-DAS~\cite{xinyi:DASDL}, hence we can focus our attention on mitigating the CCI inside a single cell, as shown in Fig.~\ref{fig:Tpgy_DAS}. We consider a \textit{symmetric network}, where every cell has the same system configuration. Without any loss of generality, we focus our attention on cell $B_0$, which is assumed to be at the origin of Fig.~\ref{fig:Tpgy_DAS}. 

In the case of the MR-FFR-DAS arrangement of Fig.~\ref{fig:Tpgy_DAS}, we assume that $N_{r}$ RAs are employed and a total of $N_{ms}$ active MSs are roaming in the cell-edge area. Additionally, in order to increase the attainable  diversity gain, in each scheduling period $N_{mr}$ half-duplex MRs are invoked for supporting our FFR-DAS. Furthermore, a single omni-directional antenna-element is employed both by each RA and by each MS. For the sake of simplicity, $N_{ms} = N_{mr} = N_t$ is assumed. The $N_t$ MRs roaming in the cell-edge area are denoted by $M_k^e$, $k \in \{1, \cdots, N_t\}$, which are identified by their polar coordinates, similarly to the actively communicating MSs, as seen in Table~\ref{table:tp_p}. 
\begin{table} [tbp]
\extrarowheight 2pt
\renewcommand{\arraystretch}{1.1}
\begin{center}
\caption{Topology Parameters for MR-FFR-DAS and CoMP-CAS Schemes.}
\label{table:tp_p}
\begin{footnotesize}
\begin{tabular}{|c|c|r|}
\hline
\hline
\textbf{MR-FFR-DAS}  & \textbf{Location} & \textbf{Polar coordinates} \\ 
\hline
\hline 
RA $R_i^e$ & cell-edge    & $(\theta_{R_i^e},L_{R_i^e}) = (\frac{2\pi(i-1)}{N_r}, d)$,\\
   &                     &  $i \in [1,\cdots, N_r]$\\
\hline
MS $Z_k^e$ & cell-edge    & $(\theta_{Z_k^e},L_{Z_k^e})$, $k \in [1, \cdots, N_t]$, \\
   &                     & roaming randomly\\
\hline
MR $M_k^e$  & cell-edge    & $(\theta_{M_k^e},L_{M_k^e})$, $k \in [1, \cdots, N_t]$,   \\
   &                     &  roaming randomly\\
\hline
\hline
\textbf{CoMP-CAS}  & \textbf{Location} & \textbf{Polar coordinates} \\ 
\hline
\hline
MS $Z_k^b$ & collaboration area  & $(\theta_{Z_k^b},L_{Z_k^b})$, $k \in [1, \cdots,  N_t]$,\\
   &        & roaming randomly\\
\hline
\hline
\end{tabular}
\end{footnotesize}
\end{center}
\end{table}

\subsubsection{CoMP-CAS Architecture}
\begin{figure}[tbp]
    \centering
    {
        \includegraphics[width=0.9\linewidth]{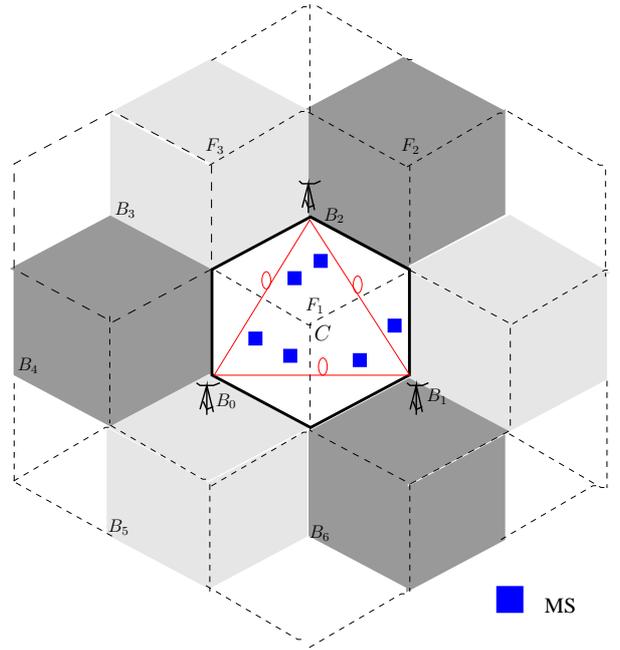}
        
    } 
    \caption{The topology of the CoMP-CAS composed by three adjacent BSs, where $N_r = 6$ collocated antennas are employed at each BS and $N_t = 6$ MSs randomly roam in the cell-edge area.}
   \label{fig:Tpgy_BSCP}
\end{figure}
The CoMP-CAS architecture considered is shown in Fig.~\ref{fig:Tpgy_BSCP}, where each single cell is divided into three $120^{\circ}$ sectors $S_i, i\in\{1, 2, 3\}$~\cite{5165405} and $N_r$ collocated antennas are employed at each BS~\cite{BScp:Marsch}. We assume that the classic frequency-division multiplexing (FDM), associated with $F_i, i\in\{1,2, 3\}$, is used for the corresponding sectors $S_i$. Hence every set of three adjacent BSs constitutes a CoMP transmission/reception area, as shown in Fig.~\ref{fig:Tpgy_BSCP}, where $B_0$ is assumed to be at the origin. We assume that there are $N_t$ active MSs in the CoMP area of Fig.~\ref{fig:Tpgy_BSCP} and all these MSs transmit at the same frequency on the uplink. Additionally, each MS employs a single omni-directional antenna-element, while roaming in the CoMP area.

The BS cooperation aided cellular network of Fig. \ref{fig:Tpgy_BSCP} is also \textit{symmetric}. Hence, without any loss of generality, we assume that the MS $Z_1^b$ roams along the line $\overline{B_0C}$, namely in the direction between $B_0$ and the center of the CoMP area, while the remaining MSs $Z_k^b, k \in \{2, \cdots, N_t\}$ randomly roam across the entire $B_0$-centered cell area according to a uniform distribution. Their polar coordinates are shown in Table~\ref{table:tp_p}.

Hence, when observing the $N_t$ active MSs, the multiple virtual MIMO channel matrices of the cell-edge area transmissions taking place in our MR-FFR-DAS scheme of Fig.~\ref{fig:Tpgy_DAS} as well as the single virtual MIMO matrix of the conventional CoMP-CAS scheme\footnote{A more detailed description of these virtual MIMO matrices is given in Section \ref{secUL_rs}.} shown in Fig.~\ref{fig:Tpgy_BSCP} has the same size of ($N_r\times N_t$)-elements. Our MR-FFR-DAS scheme gleans cooperative diversity gain from the MRs, albeit this is achieved at the cost of invoking a two-time-slot cooperation protocol. To elaborate a little further, the $N_t$ active MSs transmit during the first time slot and the corresponding $N_t$ MRs retransmit their received signal in the second time slot. By contrast, the conventional CoMP scheme gleans its cooperative diversity gain from the adjacent two BSs.\footnote{In practice, for CoMP aided systems, besides the time used by the transmission between the BS and the MSs, there is an additional delay imposed by sharing data/channel information between the collaborative BSs. More specifically, in the CoMP-CAS uplink considered, actually we also need more than one, if not two, time slot to finish a single transmission: in the first time slot the MSs transmit and in the second time slot the collaborative BSs transmit the decoded data to each other for the sake of making a final decision. Therefore, in terms of the transmission time required, the schemes considered may be regarded identical. }

\subsubsection{Digital Fibre Soliton Aided Backhaul}
\label{sec_soliBH}
\begin{figure}[t]
\centering
    \includegraphics[width=\linewidth]{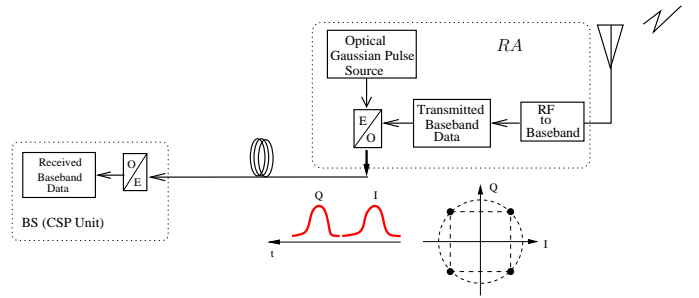} %
    \caption{System architecture of digital fibre optical link.}
    \label{fig:Fiber_soli}
 \end{figure}

Until recently the optical fibre backhaul has been assumed to be a perfect channel, when transmitting low-rate data using on-off keying (OOK). However, when aiming for supporting Gigabit-transmissions, which is the ambitious goal of LTE-A standard~\cite{xx08r:LTEA2010}, the high-rate fibre-based backhaul may suffer from the detrimental effects of both dispersion and nonlinearity~\cite{RoF:wake2010}. Since our MR-FFR-DAS scheme still relies on centralized signal processing\footnote{The distributed RAs themselves do not have computing power, but only collect or radiate signals.} at the BS, where the signals are received from the RAs via optical fibre links, these signals may be contaminated both by the fibre's dispersion and by its nonlinearity. Fortunately, these degradations may be effectively mitigated by using the fibre soliton~\cite{RoFlau, optical:Agrawal2006}. To elaborate a little further, the fibre soliton technique of \cite{RoFlau,  optical:Agrawal2006} is capable of improving both the linear and non-linear distortion of the optical fibre backhaul. As a result, the optical pulse can propagate over the optical fibre with no distortion, despite the interplay between the dispersive and nonlinear effects~\cite{optical:Agrawal2006}. Fig.~\ref{fig:Fiber_soli} shows a single optical fibre link from a RA to the BS, where QPSK modulation is applied on the uplink of the MR-FFR-DAS scheme. The signals received by the RAs from the wireless channel are first down-converted to the baseband. Then, the I- and Q-streams are modulated by optical pulses, and the resultant optical signaling pulses are transmitted through the optical fibre.

\section{Received Signal Models of the Uplink MR-FFR-DAS}
\label{secUL_rs}
\begin{figure}[t]
\centering
    \includegraphics[width=\linewidth]{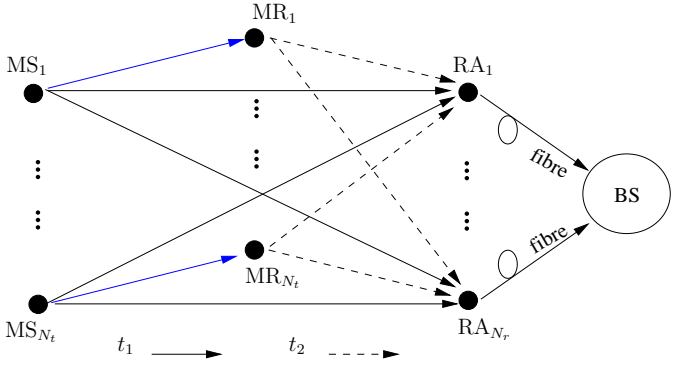} %
    \caption{Diagram of a DAS based network with $N_t$ MSs assisted by $N_t$ MRs, which is modeled as a multiuser multi-relay system. The signal received at $N_r$ RAs via the wireless channel are transmitted through the optical fibre to the BS that plays the the role of central signal processing (CSP).}
    \label{fig:multiMSMR}
\end{figure}

As shown in Section~\ref{sec2UL}, both the MR-FFR-DAS of Fig.~\ref{fig:Tpgy_DAS} and the CoMP-CAS of Fig.~\ref{fig:Tpgy_BSCP} may be modeled relying on the virtual MIMO concept. For the sake of clarity, the signal transmission process of the MR-FFR-DAS scheme considered is illustrated in Fig.~\ref{fig:multiMSMR}. We can see from Fig.~\ref{fig:multiMSMR} and Fig.~\ref{fig:Tpgy_BSCP} that both the FFR-DAS scheme using no MRs as well as the conventional CoMP-CAS may be modeled as a multi-source, multi-destination network having direct links only, where the multiple destinations are interconnected and hence they are capable of conducting collaborative signal processing. Therefore, their wireless transmission part may be modeled as a single $(N_r \times N_t)$-element virtual MIMO system. By contrast, the FFR-DAS scheme assisted by $N_t$ half-duplex MRs may be modeled as a multiuser multi-relay network having both a direct link and a two-hop relay link for each MS. We assume that each MS is served by (possibly) multiple RA elements and a single selected MR, which is located between the MS and the serving RA elements. Additionally, the serving MRs of the MSs transmitting their signals simultaneously are assumed to be sufficiently far from each other. Hence, when observing the $k$th MR in Fig.~\ref{fig:multiMSMR}, the interference from the $j$th MS, $j\neq k$, may be ignored.\footnote{Note that this assumption is indeed realistic upon invoking carefully designed MR selection, as detailed in Section \ref{sec_RLsel}. Since the MRs are single-antenna nodes and are distributed, it is infeasible for them to conduct joint detection. Thus, the virtual MIMO system model is inappropriate for the transmissions between the MSs and MRs. By contrast, we do not have to impose this assumption if more complex multi-antenna MRs are used, because in this case the multiuser joint detection technique can be employed at each MR for mitigating the impact of interference imposed by other MSs.} As a result, the wireless transmission part of the MR-FFR-DAS scheme may be modeled as a pair of $(N_r \times N_t)$-element virtual MIMO subsystems (accounting for the direct links from the MSs to the RAs as well as the links from the MRs to the RAs, respectively) and a subsystem with $N_t$ parallel single-input single-output links.  

\subsection{Received Signal of a Single Optical Fibre Link}
\label{secUL_rs_sl}
For the $i$th optical fibre link $RA_i \rightarrow BS$, $i = 1, 2, \cdots, N_r$, the soliton technique of~\cite{optical:Agrawal2006} may be applied for eliminating both the linear and non-linear distortions. As a result, a near-perfect optical fibre backhaul may be created, where the optical signaling pulses are capable of propagating without distortion, as mentioned in Section~\ref{sec_soliBH}. Hence the phase rotation imposed by the optical fibre link on the modulated signal is negligible and the modulated signal's amplitude is also maintained, albeit naturally the modulated signal is contaminated by the complex-valued additive white Gaussian noise (AWGN) $n_f \sim \mathcal{CN}(0,\sigma^2_f)$ at the BS receiver. 

As far as the wireless transmission part is concerned, let us consider the direct links between the MSs and $RA_i$ as an example. Then, the signal received at $RA_i$ from the wireless channel may be written as: 
\begin{equation}\label{eq:direct_wireless}
  r_i = \sum_{k =1}^{N_t}\zeta_{ik}h_{ik}x_k + n_w, 
\end{equation}
where $x_k$, $\zeta_{ik}$, $h_{ik}$ and $n_w \sim \mathcal{CN}(0,\sigma^2_w)$, $k =1,2, \cdots, N_t$, $i = 1, 2, \cdots, N_r$, represent the signal transmitted from $MS_k$, the pathloss between $MS_k$ and $RA_i$, the small-scale Rayleigh fading coefficients between $MS_k$ and $RA_i$, and the AWGN corresponding to the signal $r_i$ received at $RA_i$, respectively.\footnote{We assume that the noise variance of each direct wireless link  remains equal to $\sigma^2_w$.} Then, the received signal of the $i$th optical fibre link at the BS may be written as $y_i = \chi r_i +n_f$, which is expressed more explicitly as:
\begin{eqnarray}\label{eq:received_signal_ith_link}
y_i = \sum_{k = 1}^{N_t}\underbrace{\chi\zeta_{ik}}_{g_{ik}}h_{ik}x_k + \underbrace{\chi n_w + n_f}_{n_i},
\label{eq_ULrs}
\end{eqnarray}
where $\chi$ is the power-scaling factor invoked for ensuring that the peak power of the optical signaling pulse obeys the fundamental soliton requirement of~\cite{optical:Agrawal2006}. Hence $g_{ik} = \chi \zeta_{ik}$ represents the equivalent $MS_k$-$RA_i$-BS link's large-scale attenuation. Finally, $n_i = \chi n_w + n_f$ denotes the equivalent receiver noise jointly induced by the optical fibre and the wireless channel. It is also worth noting that for each $MS_k$, an obvious near-far effect is observed at the RAs. More explicitly, for each $MS_k$, the SINRs at the RAs may differ significantly, since $MS_k$ is mainly served by its nearest RA.

Additionally, when MRs are invoked, the signal received at $RA_i$ from the MRs may be characterized in a similar fashion to Eq. (\ref{eq:direct_wireless}) and (\ref{eq:received_signal_ith_link}), but with smaller pathloss, different power-scaling factor and different channel fading coefficients. 

\subsection{Received Signal Model of the Virtual MIMO Observed at the BS}
\label{secUL_rs_VM}
The channel state information (CSI) of all $MS_k$-$RA_i$-BS links is assumed to be perfectly known at the central BS. Again, let us consider the direct links from the $N_t$ MSs to the BS as an example. Based on Eq.~(\ref{eq_ULrs}), the signal vector received at the BS on the idealized synchronous uplink may be written as:
\begin{eqnarray}
{\bf y} = {\bf H}{\bf x} + {\bf n},
\label{eq_rs_vm}
\end{eqnarray}
where ${\bf y}\in \mathbb{C}^{N_r \times 1}$, ${\bf H} \in \mathbb C^{N_r\times N_t}$ and ${\bf n} \in \mathbb{C}^{N_r \times 1} $ denote the received signal vector, the channel matrix containing the perfect CSI and the noise vector, respectively. Furthermore, ${\bf x} = [x_{1}, x_{2},\cdots,x_{N_t}]^T$ represents the symbol vector transmitted from the $N_t$ MSs, while ${\bf n} = [n_{1}, n_{2},\cdots,n_{N_r}]^T$ stands for the circularly symmetric complex Gaussian \textit{effective} noise vector at the BS, and its element $n_i \sim \mathcal{CN}(0, \sigma^2)$ was defined in (\ref{eq:received_signal_ith_link}). Still referring to Eq.~(\ref{eq_rs_vm}), we have ${\bf H}=[{\bf h}_1^T,{\bf h}_2^T,\cdots,{\bf h}_{N_r}^T]^T$, where ${\bf h}_{i}=[{\tilde h}_{i1}, \tilde{h}_{i2}, \cdot, \tilde{h}_{iN_t}] \in \mathbb C^{1 \times N_t}$ represents the channel vector between all the $N_t$ MSs and the BS via a particular $RA_i$, and $ \tilde{h}_{ik}$ is defined as $g_{ik}h_{ik}$ according to Eq. (\ref{eq:received_signal_ith_link}). The entries of $\bf H$ account for both the pathloss and the small-scale Rayleigh fading of the wireless channel, as well as the impact of the optical fibre link, $i = 1,2, \cdots, N_r$, $k = 1, 2, \cdots, N_t$. 

The received signal model introduced above in Eq.~(\ref{eq_ULrs})-(\ref{eq_rs_vm}) can be extended in a straightforward manner to both the second time-slot of the MR-FFR-DAS scheme of Fig.~\ref{fig:Tpgy_DAS} and to the conventional CoMP-CAS of Fig.~\ref{fig:Tpgy_BSCP}. However, in the CoMP-CAS we have $n_i = n_w$ if a wireless backhaul is employed amongst the collaborative BSs, and typically a higher pathloss $\zeta_{ik}$ is imposed on the cell-edge MSs compared to the DAS scheme~\cite{155970}.

\subsection{Correlation Between the Channel Coefficients of $MS_k$-$RA_i$-BS and $MR_k$-$RA_i$-BS Links}
\label{sec_correlation}
We assume that the decode-and-forward (DF) relaying protocol is invoked in the multiuser MR-FFR-DAS scheme of Fig. \ref{fig:multiMSMR}. Since the MRs are geographically distributed, joint detection/decoding at the MRs is infeasible. Therefore, in the first time slot we have $N_t$ single-input single-output links and an $(N_r \times N_t)$-element virtual MIMO subsystem that relies on the signal model of Eq.~(\ref{eq_rs_vm}). More specifically, as shown in Fig. \ref{fig:multiMSMR}, the $N_t$ single-input single-output links accounts for the transmissions from the $N_t$ MSs to their selected $N_t$ MRs, and the virtual MIMO system characterizes the direct transmissions from $N_t$ MSs to the BS via the $N_r$ RAs. Furthermore, in the second time slot, the MSs remain silent when the $N_t$ MRs retransmit the decoded information to the BS, again, via the $N_r$ RAs. Hence, the channel model of the second time slot may also be regarded as an ($N_r\times N_t$)-element virtual MIMO model, similar to Eq. (\ref{eq_rs_vm}). 

Similar to the virtual MIMO channel $\bf H$ of Eq.~(\ref{eq_rs_vm}), which characterizes all the $MS_k$-$RA_i$-BS links, let us instantiate the channel matrix representing all the $MR_k$-$RA_i$-BS links as ${\bf H}_R$, where we have  
${\bf H}_R=[\hat{{\bf h}}_{1}^T,\hat{{\bf h}}_2^T,\cdots,\hat{{\bf h}}_{N_r}^T]^T \in \mathbb C^{N_r\times N_t}$. 
More explicitly, during the second time slot, for the MRs retransmitting their signals received in the fist time slot, we have $\hat{\bf{h}}_{i} = [{\hat h}_{i1}, {\hat h}_{i2}, \cdots, {\hat h}_{iN_t}]\in \mathbb{C}^{1\times N_t}$, $i\in[1,\cdots, N_r]$, which represents the channel vector from all the $N_t$ MRs to the BS via a particular $ RA_i$. 

When the $MS_k$ is close to the selected $MR_k$ and far from the other MRs,\footnote{The selected $MR_k$ exhibits the best performance among all MRs for the $MS_k$-$MR$-$RA_i$ link, and it is not necessarily the spatially nearest MR to $MS_k$ or to the corresponding $RA_i$\cite{1285365}.} the $MS_k$-RA-BS link and the $MS_k$-$MR_k$-RA-BS link may exhibit a high envelope correlation and the interference imposed by $MS_k$ on the rest of the MRs may be negligible. Additionally, in order to gain fundamental insights, in this scenario it is reasonable to assume that the selected $MR_k$ employs perfect DF relaying, implying that the source signal of the $k$th MS is perfectly decoded at the $k$th MR~\cite{1321222}, although in practice we might still observe some degree of decoding error propagation at the MR. As a result, the single-input single-output $MS_k$-$MR_k$ link becomes lossless, and only the envelope correlation between the $MR_k$-$RA_i$-BS channels ${\bf H}_R$ and the direct $MS_k$-$RA_i$-BS links' channel $\bf{H}$ is relevant. This 
envelope correlation is given by\cite{Kyritsi_2003:correlation_analysis_MIMO}
\begin{equation}
\rho_{ik} =\frac {\text E (|\tilde h_{ik}| |\hat h_{ik}| ) -\text E(|\tilde h_{ik}|) \text E(|\hat h_{ik}|) } { \sqrt{ \left(\text E( |\tilde h_{ik}|^2 ) - [\text E (|\tilde h_{ik}|)]^2  \right) \left( \text E( |\hat h_{ik}|^2 ) - [\text E (|\hat h_{ik}|)]^2 \right) }},   
\label{eq_coffRL}
\end{equation}
where $k\in[1, \cdots, N_t]$ and $ i\in[1,\cdots, N_r]$. For $\rho_{ik} = 0$ the $MS_k$-$RA_i$-BS link in the first time slot and its corresponding $MR_k$-$RA_i$-BS link in the second time slot are regarded to be uncorrelated.

Similarly, when the selected relay $MR_k$ is close to $RA_i$, we assume that the signal transmitted by $MR_k$ is perfectly received at $RA_i$ and that the interference imposed by $MR_k$ on the rest of the RAs is negligible. Hence, only the envelope correlation between the $MS_k$-$MR_k$ link and the direct $MS_k$-$RA_i$ link is relevant, which may be characterized in a similar fashion to Eq. (\ref{eq_coffRL}).

\section{Central Signal Processing at the BS}
\label{secUL_CP}
The philosophy of the DAS architecture relies on invoking RAs for transmitting (on the downlink) and receiving (on the uplink) the signals, which facilitates the centralized processing of the virtual MIMO signals at the BS, since the BS can afford to apply more complex MUD techniques. For a forward error correction (FEC)-coded system, if a soft MUD is invoked, the log-likelihood ratio (LLR) of each coded bit is calculated based on the signal vectors received at the BS. Then, the LLR of each bit is subjected to soft decoding, since soft decoding is capable of achieving a better performance than hard decoding. For the sake of maintaining a low computational complexity, in this paper we opt for an open-loop soft receiver dispensing with iterations between the MUD and the FEC decoder. Before introducing our powerful PDA-based soft MUD that is capable of achieving an attractive performance at the expense of a moderate computational complexity, let us first introduce the optimal ML based soft MUD and the classic 
MMSE-OSIC based hard-decision MUD as our benchmarkers.

\subsection{Optimal ML Based Soft MUD}
\label{sec_ML}
For an FEC-coded virtual MIMO system, the soft ML-MUD is the optimum detector, albeit it imposes a potentially excessive complexity. The soft ML-MUD calculates the LLR for the $n$th bit $b_{k,n}$ of the $k$th MS according to:
\begin{eqnarray}
 L(b_{k,n})  = \log \frac {P({\bf y}|b_{k,n} = 1)}{P({\bf y}|b_{k,n} = 0)}.
\label{eqUL_LLR} 
\end{eqnarray}

The max-log approximation may be invoked for reducing the complexity at a negligible performance degradation. Hence, the LLR in Eq.~(\ref{eqUL_LLR}) may be reformulated as \cite{5288446}
\begin{align}
 L(b_{k,n}) = &\log \left( \sum_{x_1\in {\mathbb{A}},\cdots, x_k\in {\mathbb{A} }_n^{(1)}, \cdots, x_{N_t}\in {\mathbb A}}\exp \left(-\frac{\Vert{\bf y} - {\bf   Hx}\Vert^2} {\sigma^2} \right)\right) \nonumber \\
            & - \log \left( \sum_{x_1\in {\mathbb A},\cdots, x_k\in {\mathbb A}_n^{(0)}, \cdots, x_{N_t}\in {\mathbb A}}\exp \left(-\frac{\Vert{\bf y} - {\bf Hx}\Vert^2} {\sigma^2} \right)\right) \nonumber \\
            \approx & \frac{1}{\sigma^2}\left( \min_{x_1\in {\mathbb A},\cdots, x_k\in {\mathbb A}_n^{(0)}, \cdots, x_{N_t}\in     {\mathbb A}}\Vert{\bf y} - {\bf Hx}\Vert^2 \right) \nonumber \\
             & - \frac{1}{\sigma^2}\left(  \min_{x_1\in {\mathbb A},\cdots, x_k\in {\mathbb A}_n^{(1)}, \cdots, x_{N_t}\in     {\mathbb A}}\Vert{\bf y} - {\bf Hx}\Vert^2 \right) , \nonumber \\
\label{eq_LLRE}
\end{align}
where $\mathbb{A}$ denotes the constellation alphabet, and ${\mathbb A}_n^{(b)}$ represents the set of constellation symbols whose $n$th bit is equal to $b \in \{0, 1\}$. Hence, the size of ${\mathbb A}_n^{(b)}$ is half of that of $\mathbb{A}$.  More explicitly, for calculating $L(b_{k,n})$, each of the two $\min$ operators in Eq.~(\ref{eq_LLRE}) calculates $\frac{1}{2}|\mathbb{A}|^{N_t}$ Euclidean distances (EDs) and finds the minimum of them. Therefore, in total $|\mathbb{A}|^{N_t}$ EDs are calculated, which represents a brute-force search over all possible values of the symbol vector $\bf{x}$. In other words, the size of the solution-space increases exponentially with the number of MSs (or MRs) in the MR-FFR-DAS scheme considered.

\subsection{MMSE-OSIC Based Hard-Decision MUD}
A classic reduced-complexity suboptimum MUD solution is constituted by the MMSE-OSIC detector, which is capable of striking a tradeoff between reducing the intra-cell CCI imposed by the other co-channel MSs of the same cell and mitigating the impact of the Gaussian background noise~\cite{optical:Tse} in the MR-FFR-DAS scheme considered. Relying on Eq. (\ref{eq_rs_vm}), the basic principle of the OSIC operations may be described as follows. Among all the elements of $\bf x$, the specific symbol element that has the highest receive signal-to-noise ratio (SNR), say $x_k$, was first decoded and re-modulated. Then, the impact of the corresponding regenerated transmit signal, which is the product of $x_k$ and the $k$th column vector of $\bf H$, is subtracted from the received signal vector $\bf y$. Subsequently, this process is repeated in the next interference cancellation stage relying on the residual received signal vector. Theoretically, the MMSE-OSIC detector is capable of approaching the MIMO capacity\cite{optical:Tse}. Unfortunately, in practice this optimality is undermined, since the MMSE-OSIC detector is prone to inter-layer error propagation in the presence of decision errors at each layer~\cite{optical:Tse}.

\subsection{PDA Based Soft MUD}
\label{sec_PDA}
As noted in Section~\ref{sec_ML}, the optimal ML based MUD has a computational complexity that increases exponentially with the number of MSs simultaneously served. Hence, it may not be invoked in practical cellular systems having a high number of MSs. As an attractive low-complexity design alternative, the PDA algorithm that is capable of generating bit LLRs for the concatenated channel decoder without a brute-force search may be applied in our MR-FFR-DAS scheme. In the uplink scenario considered, we assume that the CSI $\textbf{H}$ is unknown at the transmitters of the MSs, but it can be accurately estimated at the receiver of the BS. Furthermore, for the sake of convenience, $N_t = N_r$ is assumed and the decorrelated signal model\cite{Yang:BSCoop} is adopted. Hence, Eq.~(\ref{eq_rs_vm}) may be reformulated as
\begin{eqnarray}
{\bf \underline y} = {\bf x} + {\bf \underline n} = x_k{\bf e}_k + \underbrace{\sum_{j\neq k}x_j{\bf e}_j + {\bf \underline n}}_{{\bf v}_k} ,
\label{eq_rs_pda}
\end{eqnarray}
where ${\bf \underline y} = ({\bf H}^H{\bf H})^{-1}{\bf H}^H{\bf y}$, ${\bf \underline n}$ is a colored Gaussian noise vector with zero mean and covariance matrix of $N_0({\bf H}^H{\bf H})^{-1}$, ${\bf e}_k$ is a column vector with a $1$ in the $k$th position and $0$ elsewhere, while ${\bf v}_k$ denotes the interference plus noise term for symbol $x_k, k \in [1,\cdots, N_t]$. For each symbol $x_k$, we have a probability vector ${\bf p}_k$, whose $m$th element $\mathcal{P}_({x}_k = a_m|{\bf y})$ quantifies the current estimate of the \textit{nominal} \textit{a posteriori probability} (APP)\footnote{As shown in\cite{Yang:nominal_to_true_APP}, since an approximate form of the Bayes' theorem is typically invoked in the PDA algorithm, this nominal APP is essentially the normalized symbol likelihood, rather than the true APP.} of having $x_k = a_m, m\in[1,\cdots, M]$ upon receiving $\bf y$, where $a_m$ represents the $m$th element of the modulation constellation alphabet $\mathbb{A}$.

The basic idea of the PDA algorithm is to iteratively approximate the complex random vector ${\bf v}_k$ that obeys the multimodal Gaussian mixture distribution as a single multivariate colored Gaussian distributed random vector having an iteratively updated mean, covariance and pseudocovariance\footnote{The pseudocovariance is necessary for characterizing complex random vector that is improper\cite{Neeser:proper_Gaussian_process,Adali2011:complex_valued_SP, Mandic_2009:complex_valued_signal_processing}.} given by 
\begin{align}
\text E({\bf v}_k) =&\sum_{j\neq k} \text E (x_j) {\bf e}_j,  \\
{\text C}({\bf v}_k) =&\sum_{j\neq k}{\text C}(x_j){\bf e}_j{\bf e}_j^T + N_0({\bf H}^H{\bf H})^{-1},  \\
{\underline{\text C}}({\bf v}_k) =&\sum_{j\neq k} {\underline {\text C}}(x_j){\bf e}_j{\bf e}_j^T, 
\end{align}
respectively.
where we have
\begin{align}
\text E (x_j) = &\sum_{m=1}^{M}a_m\mathcal P({x_j = a_m}|{\bf y}),\\
{\text C}(x_j) = &\sum_{m=1}^{M}[a_m - \text E(x_j)][a_m - \text E(x_j)]^*\mathcal P({x_j = a_m}|{\bf y}),\\
{\underline {\text C}}(x_j) = &\sum_{m=1}^{M}[a_m -\text E(x_j)]^2 \mathcal P({x_j = a_m}|{\bf y}).
\label{eq_pda_EV}
\end{align}

For estimating $\mathcal P(x_k = a_m|{\bf y})$, it is initialized as $1/M$ based on the uniform distribution, and then it is updated at each iteration of the PDA algorithm, which is a process of gradually reducing the decision uncertainty concerning the event $x_k = a_m|{\bf y}$. Let
\begin{equation}
 {\bf w}_m^{(k)}= \underline{{\bf y}} - a_m^{(k)}{\bf e}_k - \text E({\bf v}_k), 
\label{eq_pda_2}
\end{equation}
and
\begin{equation}
 \varphi_m(x_k) \stackrel{\bigtriangleup}{=} \exp\left( - \left(\Re ({\bf w}_m^{(k)}) \atop \Im({\bf w}_m^{(k)}) \right)^T {\bf \Lambda}_k^{-1}  \left(\Re ({\bf w}_m^{(k)}) \atop \Im({\bf w}_m^{(k)}) \right)\right), 
\label{eq_pda_3}
\end{equation}
where the composite covariance matrix of ${\bf v}_k$ is given by
\begin{equation}
 {\bf \Lambda}_k \stackrel{\bigtriangleup}{=} \left({ \Re [{\text C}({\bf v}_k) + {\underline{\text C}}({\bf v}_k)] \atop \Im[{\text C}({\bf v}_k) + {\underline{\text C}}({\bf v}_k)]} \qquad { -\Im[{\text C}({\bf v}_k) - {\underline{\text C}}({\bf v}_k) ] \atop  \Re[{\text C}({\bf v}_k) - {\underline{\text C}}({\bf v}_k) ]} \right) 
\label{eq_pad_4}
\end{equation}
and $a_m^{k}$ indicates that $a_m$ is assigned to $x_k$, while $\Re(\cdot)$ and $\Im(\cdot)$ represent the real and imaginary parts of a complex variable, respectively.

Since no external source of the \textit{a priori} probability $P(x_k = a_m)$ is available, the decision probability of the event $x_k = a_m|{\bf y}$ is approximated as
\begin{align}
\mathcal P(x_k =  a_m\vert {\bf y}) = & \frac{p({\bf y}|x_k = a_m)P(x_k = a_m)}{\sum_{m=1}^{M} p({\bf y}|x_k = a_m)P(x_k = a_m)} \nonumber \\ 
\approx & \frac{\varphi_m(x_k)}{\sum_{m=1}^{M}\varphi_m(x_k)}.
\end{align}
Hence, an updated value of $\mathcal P (x_k = a_m | {\bf y})$ has been obtained. Based on this updated value, the above decision-probability-estimation process is repeated until $\mathcal P (x_k = a_m | {\bf y})$ has converged for all values of $k$ and $m$. Then, the bit LLR $L(b_{k,n})$ delivered to the channel decoder is given by\cite{Yang:nominal_to_true_APP}
\begin{equation}\label{eq:pseudo_symbol_APP_based_L_e}
 L(b_{k,n}|{\bf{y}}) = \ln\frac{\sum\limits_{\forall a_m\in
\mathbb{A}_n^{(1)}}\mathcal{P}(x_k = a_m|{\bf{y}})}{\sum\limits_{\forall a_m\in
\mathbb{A}_n^{(0)}}\mathcal{P}(x_k =
  a_m|{\bf{y}})}.
\end{equation}
For further details on the PDA-aided MUD, please refer to~\cite{Yang:BSCoop, Yang:nominal_to_true_APP}.

\subsection{Combining the Soft Information of the MSs and MRs at the BS}
When using channel codes that support soft-input--soft-output decoding (such as convolutional codes, turbo codes and LDPC codes), the outputs of the channel decoder are also bit LLRs. In order to achieve a higher diversity gain in the MR-FFR-DAS scheme considered, the soft decision information  gleaned from the MSs and MRs may be combined in a manner similar to that proposed for the BS cooperation aided CoMP-CAS of~\cite{Yang:BSCoop}. However, channel codes were not considered in \cite{Yang:BSCoop}, hence soft information combining was implemented in the probability domain in \cite{Yang:BSCoop}. By contrast, since each active MS is assisted by an appropriately selected MR employing the DF protocol in our channel-coded MR-FFR-DAS scheme, the channel decoders of the MRs also generate bit LLRs, which may be forwarded to the RAs. Therefore, the soft information combining in our channel-coded MR-FFR-DAS scheme has to be implemented in the LLR domain.

More specifically, if no channel codes are employed, the decision probabilities $\mathcal{P}(x_k = a_m|{\bf y}_{MS})$ calculated at the BS based on the direct transmission during the first time slot and $\mathcal {P}(x_k = a_m|{\bf y}_{MR})$ calculated relying on the MR-aided transmission in the second time slot are combined as~\cite{Yang:BSCoop}
\begin{equation}
 \mathcal{P}(x_k = a_m|{\bf y}_{C}) =  \frac{\mathcal{P}(x_k = a_m |{\bf y}_{MS})\mathcal{P}(x_k = a_m|{\bf y}_{MR})}{\sum_{m}\mathcal{P}(x_k = a_m|{\bf y}_{MS})\mathcal{P}(x_k = a_m|{\bf y}_{MR})}
\label{eq_pdaMR}
\end{equation}  
for $k\in[1,\cdots, N_t]$. Then, the BS's MUD makes a decision for each transmitted symbol $x_k$, yielding $\hat{x}_k = a_{m^{'}}$, where we have:
\begin{equation}
  m^{'} = \arg \max_{m = 1,2,\cdots M} \left\lbrace \mathcal{P}(x_k = a_m|{\bf y}_{C}) \right\rbrace. 
\end{equation}
By contrast, when soft-decoded channel codes are invoked, the soft information generated by the MUD of the BS from the direct transmission during the first time slot and from the MR-aided relay transmission during the second time slot may be simply combined as
\begin{equation}
L_C(b_{k,n})  = L_{MS}(b_{k,n}) + L_{MR} (b_{k,n}).  
\end{equation}
Then, $L_C(b_{k,n})$ is fed into the channel decoder to generate the final decoding results.

\subsection{Power Control in the Multiuser Uplink Scenario}
Perfect CSI is typically unavailable in practice, whilst having an imperfect CSI leads to a degraded performance. Hence, we introduce power control for improving the SINR of the MSs roaming in the cell-edge area. We use the normalized signal-to-interference ratio (SIR) based model for investigating the effect of the multiuser interference in the uplink, since the AWGNs of both the wireless and the optical fibre links are moderate. In other words, in the multiuser multicell scenario considered, it is the effect of the interference, rather than the AWGN, that dominates the attainable performance. This is the so-called interference-limited scenario. More explicitly, in interference-limited systems we have $\mathsf{SINR}\approx \mathsf{SIR}$ and we record the SIR for all MSs randomly roaming across the cell-edge area. Using no power control, we adopt the theoretical SIR model that takes no account of fading of any kind and is solely determined by the path-loss. This simplified SIR model is expressed as~\cite[
Chapter 7]{video2001}:
\begin{align}
\mathsf{SIR}_i(\text{dB}) & = 37.6\log_{10}\frac{d_i}{d_{w}}, \label{eq_distances}\\
\mathsf{SIR}_{MS}(\text{dB}) & = -10\log_{10}\left[\sum_{i=1,\cdots, N_t -1} 10^{-\frac{\mathsf{SIR}_i(\text{dB})}{10}}\right],
\label{eq_sir_ul}
\end{align}
where $\mathsf{SIR}_i$ represents the SIR experienced by the wanted MS when there is only a single interferer having the index $i$. More specifically, $d_{w}$ is the distance between the wanted MS and its serving RA, while $d_i$ is the distance between the single interfering MS and the RA that serves the wanted MS. We assume a $37.6$dB/decade inverse power path-loss law. For the multiuser scenario considered in Fig.~\ref{fig:Tpgy_DAS}, we define $\mathsf{SIR}_{MS}$ as the SIR recorded at its respective serving RA for any MS roaming across the cell-edge area. Since a specific MS suffers from the contaminating effects of $(N_t -1)$ interferers, $\mathsf{SIR}_{MS}$ can be written in dBs as Eq.~(\ref{eq_sir_ul}).

The geographic SIR distribution of the cell-edge area MSs is based on Eq.~(\ref{eq_sir_ul}), where the specific MSs suffering from a lower SIR will be assisted by increasing their transmit power. When applying power control, the SIR of the wanted MS contaminated by the $i$th interfering MS may be written as:
\begin{eqnarray}
 \mathsf{SIR}_i^{'}(\text{dB}) = 37.6\log_{10}\frac{d_i}{d_w} + P_{dB}^{w} - P_{dB}^{i},
\label{eq_sir_pc}
\end{eqnarray}
where $d_i$ and $d_w$ were defined in Eq.~(\ref{eq_distances}). Furthermore, $P_{dB}^{w} = 10\log_{10} P^{T}_w$ and $P_{dB}^{i} = 10\log_{10} P^{T}_i$ represent the transmit power of the wanted MS and that of the interfering users in dB, respectively. More explicitly, when the MS roaming in the cell-edge area suffers from a lower $\mathsf{SIR}$, the transmit power $P_{dB}^w$ (or $P_{dB}^i$) will be increased (or decreased) in the interest of maintaining the target SIR of this MS. Correspondingly, the interfering MS may also suffer from a lower $\mathsf{SIR}$ and hence its transmit power $P_{dB}^i$ will be increased, which will in turn increase the interference imposed on the other MSs.

The expression of the SIR recorded in the presence of power control for our multiuser scenario remains similar to that without power control. Hence, based on Eq.~(\ref{eq_sir_ul}) and Eq.~(\ref{eq_sir_pc}), we have 
\begin{align}
\mathsf{SIR}_{MS}^{'}(\text{dB}) = -10\log_{10}\left[\sum_{i=1, \cdots, N_t-1} 10^{-\frac{\mathsf{SIR}_i^{'}(\text{dB})}{10}}\right].
\label{eq_sirdB_pc}
\end{align}

\section{Performance Evaluation}
\label{secUL_PfEv} 
In this section, both the uplink BER and the \textit{effective throughput} of the conventional non-cooperative CAS, of the BS-cooperation aided CoMP-CAS (illustrated in Fig.~\ref{fig:Tpgy_BSCP}), of the non-cooperative FFR-DAS dispensing with MRs and of the MR-FFR-DAS (illustrated in Fig. \ref{fig:Tpgy_DAS}) are characterized. The simulation parameters are summarized in Table~\ref{tb_sp_ul}, where $d_0$ is the distance between any transmitter-and-receiver pair in kilometers. We define the cell-edge region as the area outside the radius of $r = 0.5R$, hence by definition the cell-center area is within the radius of  $r = 0.5R$, as seen in Fig.~\ref{fig:Tpgy_DAS}. The RAs in the cell-edge area are assumed to be located at $d_e = 0.7R$. Four types of MUDs, i.e. the high-complexity non-cooperative soft ML, the non-cooperative MMSE-OSIC, the non-cooperative PDA and the cooperation-based SC-PDA, are compared in the context of both the CAS-based and FFR-DAS-based architectures, as seen in Fig.~\ref{fig:BER_BScp} and Fig.~\ref{fig:BER_best_direction} -- Fig.~\ref{fig:MR_selection_reliable_area_worst_direction}. More specifically, as a benchmarker, the classic non-cooperative MMSE-OSIC based MUD considered generates hard-decision information for the concatenated channel decoder. Hence the BER performance of the MMSE-OSIC receiver remains poor in both the CAS-based and FFR-DAS-based systems. By contrast, the soft-decision based MUDs (i.e. the soft ML, the PDA and the SC-PDA) achieve a superior BER performance, despite their increased complexity.
\begin{table}
  \caption{Simulation parameters for the uplink scenarios considered, which rely on the urban-macro propagation model of the 3GPP-LTE standard~\cite{xx08r:3gpp}.}
\label{tb_sp_ul}
  \centering
\extrarowheight 2pt
\renewcommand{\arraystretch}{1.0}
\footnotesize
 \begin{tabular}[c]{|l|r|}
\hline 
\hline 
       Urban macro BS-to-BS distance & $\overline{B_{2}B_{3}} = 3$ km \\
\hline Cell radius  & $R = \frac{\overline{B_{2}B_{3}}}{\sqrt 3}$ \\
\hline
       Pathloss (expressed in dB) & $128.1+37.6\log_{10}(d_0)$ \\
\hline
       Transmit power of MS or MR & $[20, 30]$ dBm\\
\hline
       Noise power spectral density at RAs & $-174$ dBm/Hz\\
\hline
       Shadowing standard deviation & $\sigma_{s} = 8$ dB \\
\hline
       Normalized optical fibre link SNR & 50 dB \\
\hline
       Distance between the BS and MSs  & $d/R = (0, 1]$ \\
to the cell radius ratio &  \\
\hline Distance between the BS and each  & $d_{e} = 0.7R$\\
        RA in the cell-edge area & \\
\hline
       Length of the optical fibre & $L = 5d$\\
\hline
       Number of RAs in cell-edge area  & $N_r = 6$ \\
\hline
       Number of MSs simultaneously   & $N_t = 6$\\
	served in cell-edge area & \\
\hline
       Modulation scheme & QPSK\\
\hline
       Bit-to-symbol mapping & Gray\\
\hline
       Channel code &  punctured convolutional code \\ 
&  with coding rate of ${R_c} = \frac{2}{3}$\\
\hline
       Code constraint length &  7\\
\hline
       Code generator   &      [171, 133]\\
\hline
       Channel decoder & Viterbi algorithm\\
\hline
       MUD & soft ML, MMSE-OSIC, \\
&  PDA and SC-PDA \\
\hline
       Channel model & flat Rayleigh fading\\ 
\hline Total channel bandwidth & $B$ = 20 MHz \\
\hline Packet length & $L_p$ = 1024 bits \\
\hline Bits per modulation symbol & $M_b = \log_2M = 2$ \\  
\hline Subcarrier spacing & $B_{\text{sc}} = 15$ kHz \\
\hline Symbol rate per subcarrier & $ R_s = 15$k Baud \\
\hline Number of subcarriers & $N_{\text{sc}}$  = 1200 \\           
\hline 	
\hline
\end{tabular} 
\end{table}

\subsection{Calculation of the Effective Throughput}
We define the \textit{effective throughput} in terms of bits/second/Hz as follows:
\begin{equation}
\mathsf{C}_\text{eff} = \mathsf{C}_\text{raw} \times (1-\mathsf{BER})^{L_p},  
\end{equation}
where the packet length $L_p$ is set to $1024$ bits in our evaluations\footnote{A packet loss event happens whenever any of the bits contained in this packet is erroneously decoded at the receiver. The packet length does not change the general insights and conclusions drawn from our effective throughput comparisons.}. We assume that $N_t$ MSs transmit over the whole channel bandwidth simultaneously, hence the raw throughput $\mathsf{C}_\text{raw}$ is given by 
\begin{equation}
 \mathsf{C}_\text{raw} = \frac{N_t \times  R_c \times M_b  \times R_s \times N_{\text{sc}}} {B}. 
\end{equation}
The specific parameter values invoked for calculating $\mathsf{C}_\text{raw}$ are given in Table \ref{tb_sp_ul}. 

\subsection{BER Performance of the Conventional CAS and the BS-Cooperation Aided CoMP-CAS}
\begin{figure}[tbp]
\centering
    \includegraphics[width=\linewidth]{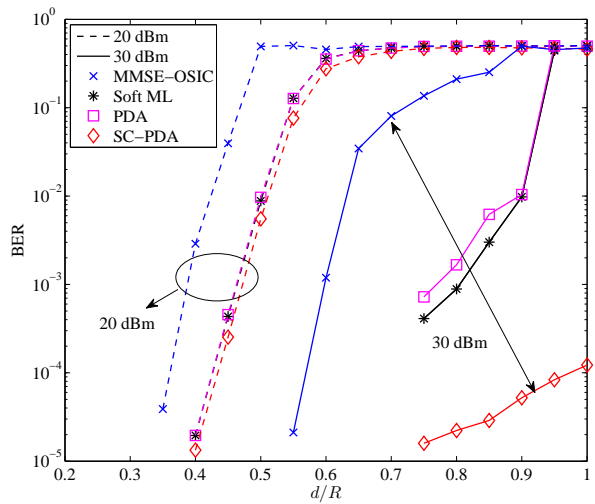} %
     \caption{BER performance of the conventional CAS architecture using the non-cooperative MMSE-OSIC, PDA and soft ML based MUDs and of the BS-cooperation aided CoMP-CAS architecture employing the SC-PDA based MUD.}
     \label{fig:BER_BScp}
\end{figure}

The BER performance of both the conventional CAS and the BS-cooperation aided CoMP-CAS schemes is characterized in Fig.~\ref{fig:BER_BScp}, which is obtained by considering $N_t = 6$ co-channel MSs roaming randomly across a single cell and across the entire collaboration area composed of three sectors of three adjacent BSs, respectively, as seen in Fig.~\ref{fig:Tpgy_BSCP}. Furthermore, $N_r = 6$ omni-directional antennas were applied at each of the BSs. Hence, in each cell there are $6$ BS antennas serving each sector, and for the CoMP-CAS, in total there are $18$ BS antennas serving the three collaborative sectors. As a result, the CoMP-CAS scheme relies on an equivalent ($18\times 6$)-element virtual MIMO system model, if perfect BS cooperation\footnote{In perfect BS cooperation, the information to be shared amongst the collaborative BSs is received at each BS without transmission error.} is assumed. When the MS/MR transmit power $P_t$ is increased from  $20$ dBm to $30$ dBm, the BERs of all MUDs considered are improved. However, this BER improvement becomes more substantial when the MSs are close to the BS, as indicated by smaller values of $d/R$. By contrast, when the MSs are roaming in the cell-edge area, as indicated by larger values of $d/R$, even an increased transmit power $P_t$ fails to result in an sufficiently competitive BER for the MS of interest. This is because the pathloss is rather high and the SINR is low in the cell-edge area. 

It is worth noting that the SC-PDA based MUD is adopted for the CoMP-CAS scheme, which requires the cooperation of three adjacent BSs. Observe from Fig.~\ref{fig:BER_BScp} that the cooperation diversity gain attained by the SC-PDA based MUD becomes more significant when the transmit power of each MS is as high as $P_t = 30$ dBm. By contrast, when the MSs transmit at a lower power of $P_t = 20$ dBm, the SC-PDA based MUD has a similarly poor BER performance to that of the non-cooperative single-cell PDA based MUD, which operates without exchanging soft information among the adjacent BSs. The reasons behind these observations are as follows. When the MSs roam close to their own anchor-BS, but quite far from other adjacent BSs, the cooperation gain remains rather limited, since the pathloss experienced by the MS of interest with regard to the adjacent BSs is high. This leads to inefficient cooperative BS processing. On the other hand, when the MSs roam close to the cell-edge, a sufficiently high SINR may only be guaranteed for a high transmit power, since the pathloss of the MSs with regard both to their anchor-BS and to any of the adjacent BSs remains high. 

\subsection{The Performance of the Non-Cooperative FFR-DAS and the MR-FFR-DAS in the Uplink}
Similarly, we consider $N_t = 6$ cell-edge MSs and $N_r = 6$ RAs in the non-cooperative FFR-DAS, while $N_t = 6$ additional MRs are also invoked in the MR-FFR-DAS scheme, as illustrated in Fig.~\ref{fig:Tpgy_DAS} and Fig.~\ref{fig:multiMSMR}. On the one hand, when no MRs are invoked, a particular MS's uplink signal received at the BS is contaminated by the other co-channel MSs transmitting within the single time slot of the non-cooperative FFR-DAS scheme. In this scenario, the performance of three non-cooperative MUDs, namely of the high-complexity soft ML, of the MMSE-OSIC and of the PDA, are numerically evaluated. On the other hand, when $N_t = 6$ MRs are employed, the signal relayed by a particular MR is also contaminated by the other co-channel MRs transmitting in the second time slot. In this context, an SC-PDA based MUD capable of exploiting the user cooperation gain gleaned from MRs is examined. In contrast to the conventional BS-cooperation aided CoMP-CAS scheme, the MR-FFR-DAS scheme attains a beneficial cooperation diversity gain with the assistance of MRs, which is 
facilitated by using a cooperation time slot as well. Furthermore, this cooperation diversity gain highly depends on the locations of the MSs. Among them, two specific directions, as represented by $\theta_{best}$ and $\theta_{worst}$, are of particular interest, since they serve as the bounds for the general case.  
\begin{figure}
\centering
\includegraphics[width=\linewidth]{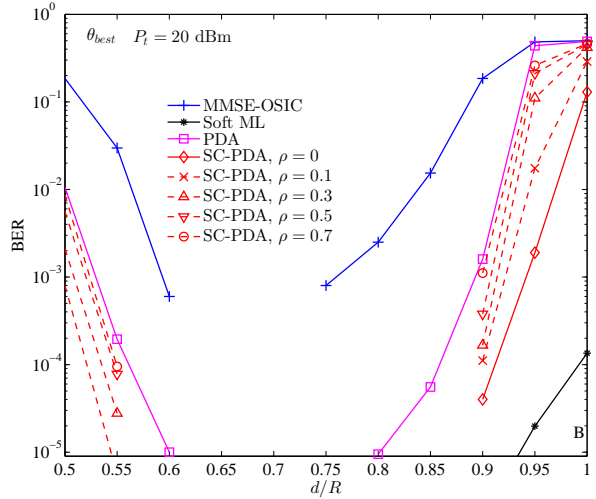}
\caption{BER performance of the MSs located at the $\theta_{best}$ direction in the cell-edge area of the non-cooperative FFR-DAS and the MR-FFR-DAS schemes. No power control is used, and a fixed MS/MR transmission power $P_t = 20$ dBm is assumed. The non-cooperative high-complexity soft ML, the MMSE-OSIC and the PDA based MUDs are invoked when MRs are not used, while the SC-PDA is used when MSs are assisted by the MRs. The MRs are selected according to the simple ``close-to-MS'' strategy.}
\label{fig:BER_best_direction} 
\end{figure}
\begin{figure}[tbp]
\centering
\includegraphics[width=\linewidth]{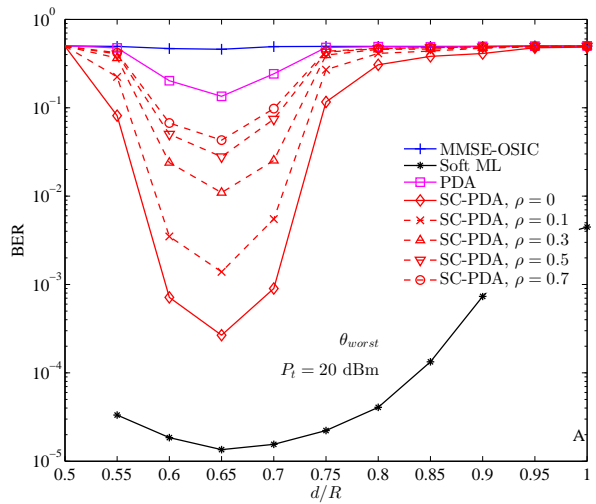}
\caption{BER performance of the MSs located at the $\theta_{worst}$ direction in the cell-edge area of the non-cooperative FFR-DAS and the MR-FFR-DAS schemes. The other configurations are the same as those of Fig. \ref{fig:BER_best_direction}.}
\label{fig:BER_worst_direction}  
\end{figure}

\subsubsection{Cell-Edge Area Performance without Power Control}
\label{sec_res_cE}
In Fig.~\ref{fig:BER_best_direction} we investigate the BER performance of both the non-cooperative FFR-DAS and the MR-FFR-DAS schemes, where the MSs are located at the $\theta_{best}$ direction in the cell-edge area, as shown in Fig.~\ref{fig:Tpgy_DAS}. No power control technique is used, and a fixed MS/MR transmission power of $P_t = 20$ dBm is assumed. Hence, for a particular MS, upon assuming a constant noise power and interfering MSs' locations, the $\mathsf{SINR}$ attained at the BS is mainly determined by the distance from this MS to the RA, as demonstrated in Eq.~(\ref{eq_sir_ul}). We can see from Fig.~\ref{fig:BER_best_direction} that when the MSs roam close to the RAs, which are at the location of $d/R = 0.7$, a high SINR may be obtained. Therefore, even the non-cooperative hard-decision MMSE-OSIC based MUD is capable of achieving a low BER. However, when the MSs roam far away from the RAs, namely towards either $d/R = 0.5$ or $d/R = 1$, the achievable SINR gradually becomes lower. As a result, both the non-cooperative MMSE-OSIC and the non-cooperative PDA fail to achieve an appealing BER in the non-cooperative FFR-DAS. By contrast, when each MS is assisted by an MR, which is selected according to the \textit{simple strategy} that the one selected has to be reasonably close to the MS for the sake of attaining a diversity gain in the MR-FFR-DAS, the SC-PDA based MUD is invoked. Since the SC-PDA is capable of efficiently combining the soft information gleaned from the MSs and MRs, it attains a BER typically lower than  $10^{-5}$ in \textit{most} locations of the cell-edge area, as defined by $d/R \in[0.5, 0.9]$. Note, however, that when the MSs are very close to the cell boundary, as indicated by $d/R \rightarrow 1$, all of the schemes considered exhibit poor BER performance.
\begin{figure*}[t]
\centering
\subfigure[]{
\label{fig:BER_improved_MR_selection_best_direction}
\includegraphics[width=0.48\linewidth]{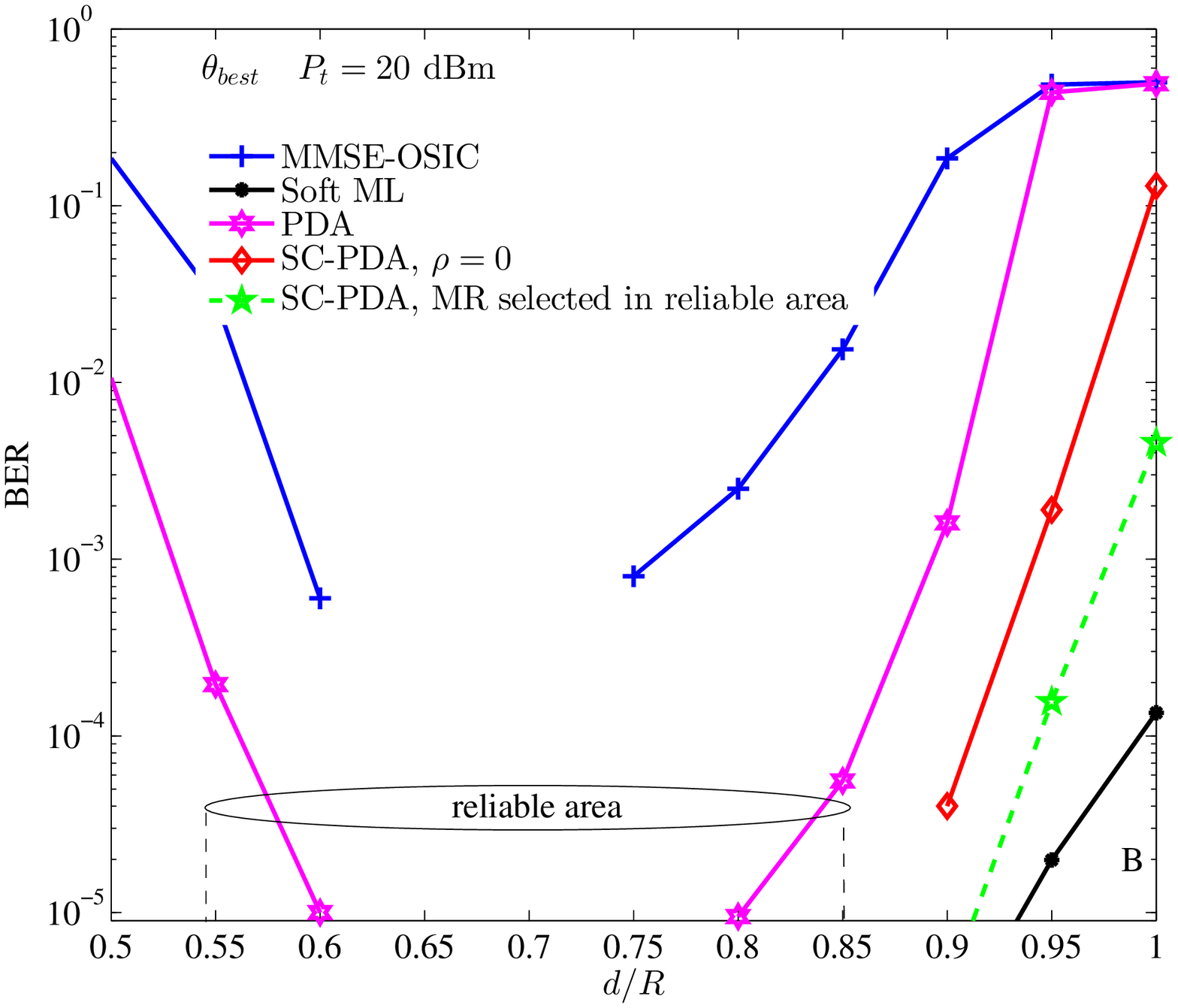}
}
\centering
\subfigure[]{
\label{fig:throughput_improved_MR_selection_best_direction}
\includegraphics[width=0.48\linewidth]{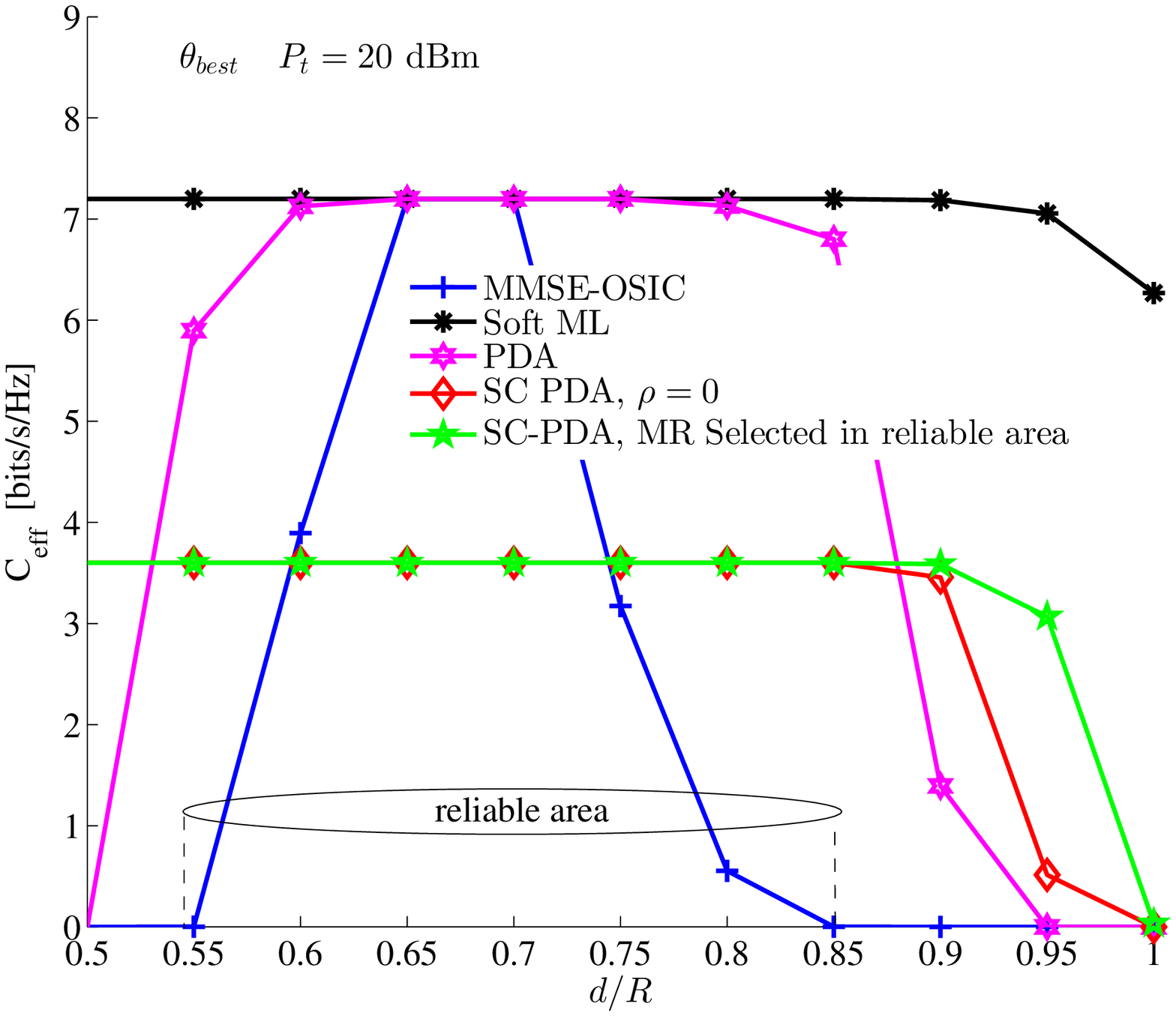}
}
\caption{Impact of the improved MR selection strategy on (a) the BER and (b) the effective throughput of the MSs located at the $\theta_{best}$ direction in the cell-edge area of the MR-FFR-DAS scheme. The ``SC-PDA, $\rho = 0$'' curves are obtained by selecting the MRs according to the simple ``close-to-MS'' strategy, which serves as a benchmark of the improved strategy that selects MRs only from the ``reliable area''. The remaining configurations are the same as those of Fig. \ref{fig:BER_best_direction}.}
\label{fig:MR_selection_reliable_area_best_direction}
\end{figure*}   

Additionally, the performance of the MSs at the $\theta_{worst}$ direction is characterized in Fig.~\ref{fig:BER_worst_direction}. In contrast to Fig.~\ref{fig:BER_best_direction}, we observe that the BER performance of even the SC-PDA based MUD is also dramatically degraded. This is because in the $\theta_{worst}$ direction, $MS_k$ roams in the area where the desired signal received at $RA_k$ may in fact be weaker than the interference imposed by other co-channel MSs. 

Therefore, it is straightforward to infer that the BER performance across the entire cell-edge area is between that of the best-case angle $\theta_{best}$, as characterized in Fig.~\ref{fig:BER_best_direction}, and that of the worst-case scenario $\theta_{worst}$, as shown in Fig.~\ref{fig:BER_worst_direction}. 
                                                                                                                                                                                                                                                                                                                                                                                                                                                                                                                                                                                                                                                                                                                                                                                                                                                                                           
Furthermore, the dotted curves seen in Fig.~\ref{fig:BER_best_direction} and Fig.~\ref{fig:BER_worst_direction} quantify the impact of the correlation between the channel coefficients of $MS_k$-$RA_k$-$BS$ and $MR_k$-$RA_k$-$BS$, as defined in Eq.~(\ref{eq_coffRL}), when the MR is selected in the vicinity of the MS. The range of the correlation coefficient $\rho$ examined is between $0$ and $1$. More explicitly, when the direct $MS_k$-$RA_k$-$BS$ link and the corresponding $MR_k$-$RA_k$-$BS$ link are uncorrelated, i.e. we have $\rho = 0$, the SC-PDA based MUD achieves a diversity gain contributed by a pair of uncorrelated channel matrices $\bf H$ and ${\bf H}_R$. By contrast, when the correlation is approaching $\rho = 1$, the diversity tends to be completely eroded, hence in this case the SC-PDA has a similar BER performance to that of the non-cooperative PDA.

\subsubsection{Improved MR Selection in the Reliable Area}
\label{sec_RLsel}
For a particular cell-edge MS and its corresponding RA, the MR selected has a location between them. In Fig.~\ref{fig:BER_best_direction} and Fig.~\ref{fig:BER_worst_direction}, when a small $\rho$ is required for obtaining an increased diversity gain, the spacing between the MR selected and the cell-edge MS has to be sufficiently large, albeit the MR is still selected in the vicinity of the cell-edge MS considered\footnote{The ``vicinity'' of a cell-edge MS may be defined as a half-circle region, which is centred at the cell-edge MS and located between this cell-edge MS and its nearest RA. When both the cell-edge MSs and the MRs are uniformly randomly distributed in the cell, it is unlikely that the MRs selected for different cell-edge MSs will be very close to each other.} rather than in the vicinity of the RA or the BS. More explicitly, the distance between the MR and the cell-edge MS should remain sufficiently large so that the correlation between the direct link and the second hop of the relayed link remains low. In this case, the SC-PDA based MUD effectively combines the soft information gleaned from the signals transmitted by both the MSs and the MRs, and a substantially improved BER may be achieved by the SC-PDA compared to both the non-cooperative PDA and the non-cooperative MMSE-OSIC. However, there is still an area where the BER cannot be significantly reduced even when invoking the SC-PDA. This is encountered when the MS roams very far away from the RA, as indicated by $d/R \rightarrow 1$. More explicitly, when the MR is selected to be close to the MS that approaches the cell-edge boundary,  both of them may be too far from the RA. As a result, the selected MRs also suffer from a high level of pathloss and the SC-PDA remains unable to effectively reduce the BER. Therefore, as far as fixed RAs are considered, for the sake of maintaining an adequate performance, the cell-edge MSs should be neither too close to the BS, nor too close to the cell-edge boundary.   

\begin{figure*}[t]
\centering
\subfigure[]{
\label{fig:BER_improved_MR_selection_worst_direction}
\includegraphics[width=0.48\linewidth]{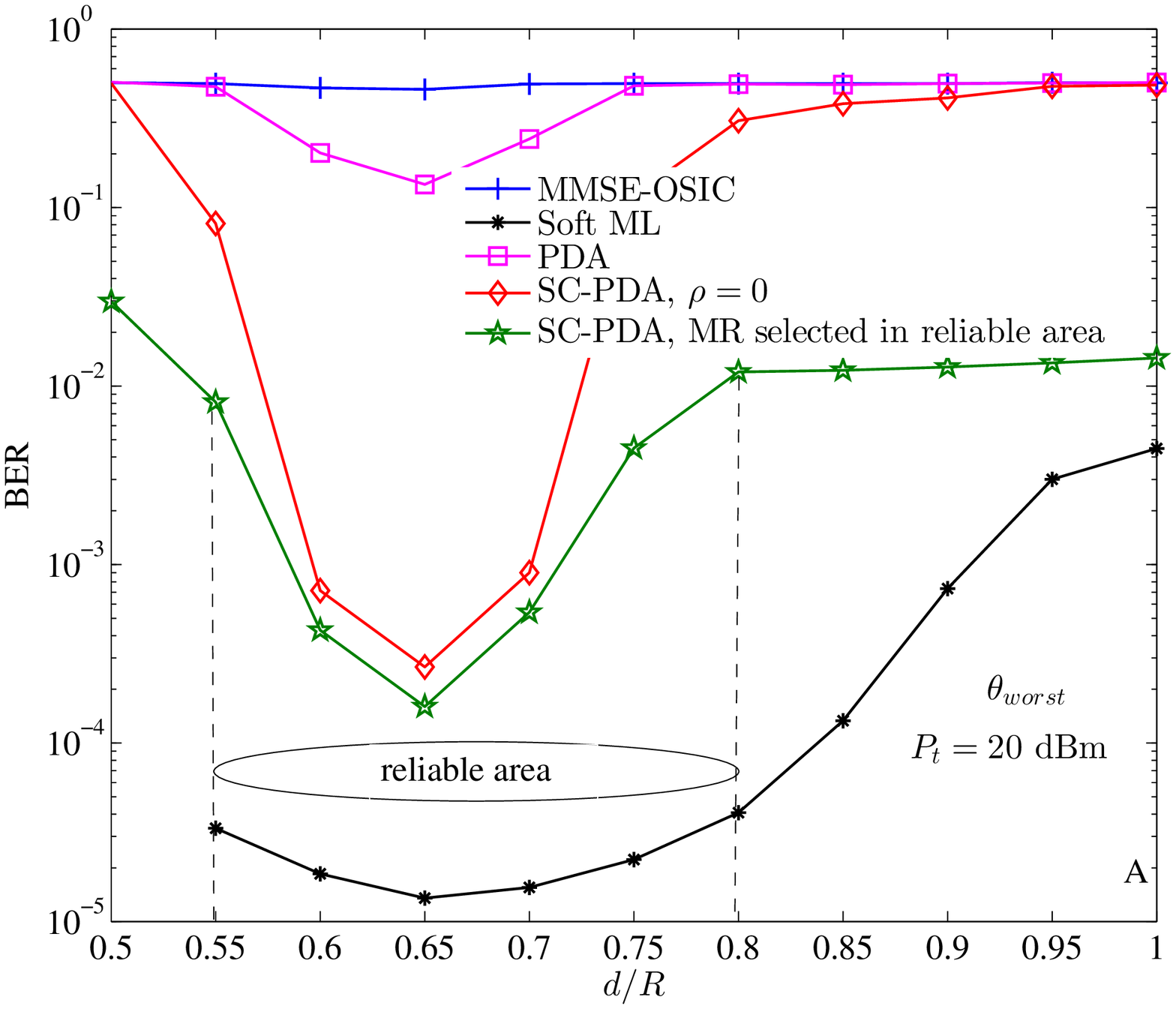}
}
\centering
\subfigure[]{
\label{fig:throughput_improved_MR_selection_worst_direction}
\includegraphics[width=0.48\linewidth]{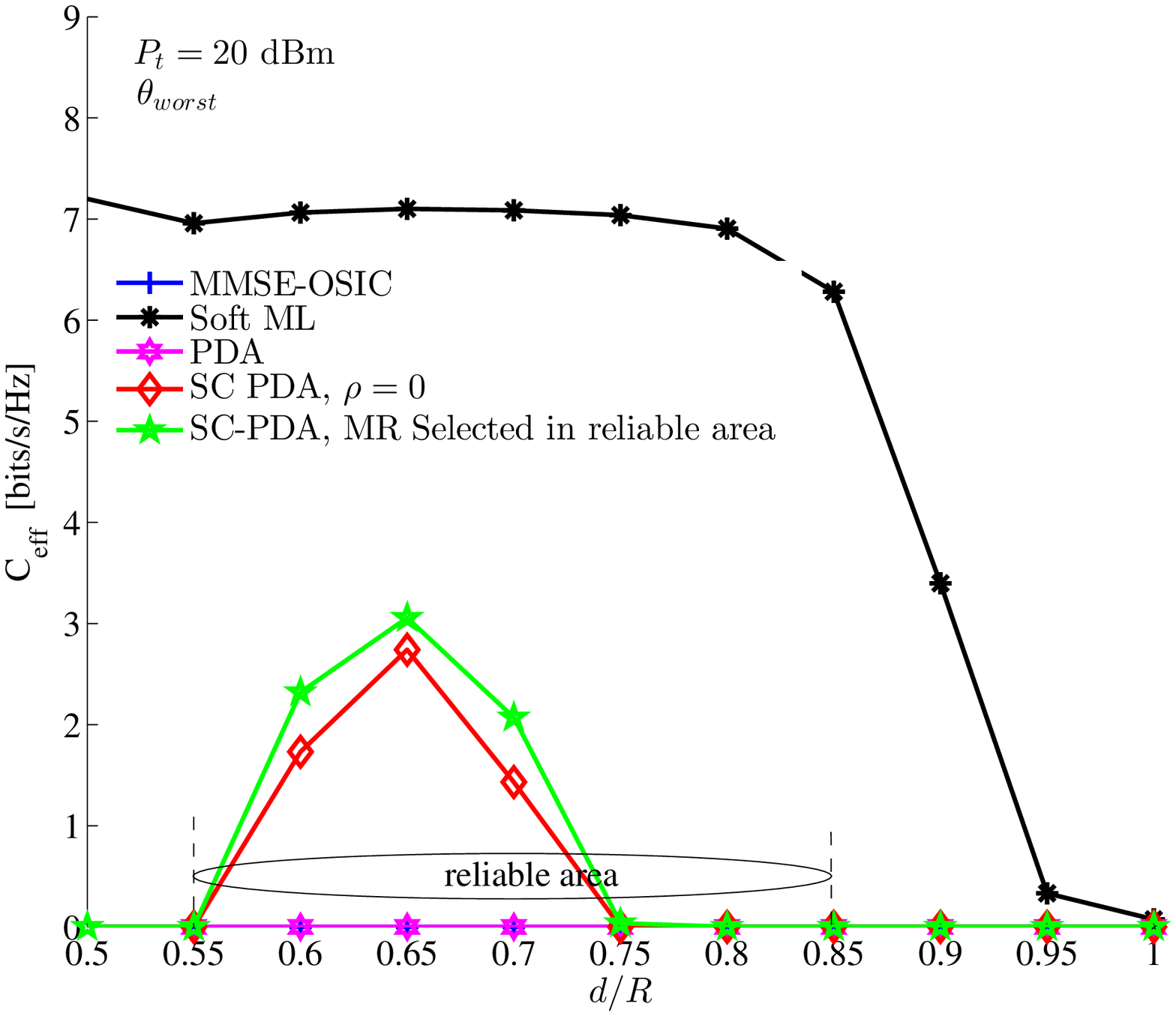}
}
\caption{Impact of the improved MR selection strategy on (a) the BER and (b) the effective throughput of the MSs located at the $\theta_{worst}$ direction in the cell-edge area of the MR-FFR-DAS scheme. The remaining configurations are the same as those of Fig.~ \ref{fig:MR_selection_reliable_area_best_direction}. }
\label{fig:MR_selection_reliable_area_worst_direction} 
\end{figure*}

As a solution to this predicament, we may always select the MR from the ``reliable area'' so that the MR selected is not too far from the RA, regardless of the ordinary cell-edge MS or the MS that is very close to the cell-edge boundary, as illustrated in Fig.~\ref{fig:MR_selection_reliable_area_best_direction} and Fig.~\ref{fig:MR_selection_reliable_area_worst_direction} for both the $\theta_{best}$ and $\theta_{worst}$ directions, respectively. Theoretically, the optimal position of the selected MR under the DF protocol should strike an attractive tradeoff between being close to the cell-edge MS and being not too far from the RA for the sake of having a good performance at both the first hop and the second hop. As demonstrated in Fig.~\ref{fig:BER_improved_MR_selection_best_direction}, as far as the SC-PDA is concerned, compared to the simple ``close-to-MS'' MR selection strategy described in Section \ref{sec_res_cE}, the original high BER experienced by the MSs when they roam far away from the RAs but remain near the $\theta_{best}$ direction may be improved by using this improved MR selection strategy. This is evidenced by the curve marked by the hollow stars in Fig.~\ref{fig:BER_improved_MR_selection_best_direction}. On the other hand, Fig.~\ref{fig:throughput_improved_MR_selection_best_direction} shows that for cell-edge MSs that are near to the RAs, the SC-PDA based MR-FFR-DAS scheme which requires two time-slots for completing each MS's transmission has a lower effective throughput than the non-cooperative FFR-DAS that uses only a single time-slot for completing the same task. However, when the MSs approach the cell-edge boundary, the MR-FFR-DAS provides a better effective throughput. This observation confirms the throughput-erosion impact of the extra time-slot imposed on the effective throughput in cooperative systems. It also demonstrates that introducing MRs is indeed particularly beneficial for the MSs roaming close to the cell-edge boundary in terms of both the BER and the effective throughput. Additionally, the benefits of the improved MR selection strategy are corroborated by the effective throughput results of Fig.~\ref{fig:throughput_improved_MR_selection_best_direction} as well. 

Similarly, we can see from Fig.~\ref{fig:BER_improved_MR_selection_worst_direction} that the improved MR selection strategy is also beneficial for the $\theta_{worst}$ direction in terms of BER. However, the performance gain is not as high as that of the $\theta_{best}$ direction. This is because the selected MRs employing the DF protocol have to be closer to the source than to the destination, and compared to the scenario of the $\theta_{best}$ direction having the same value of $d/R$, the cell-edge MSs are farther away from the RAs in the $\theta_{worst}$ direction. As a result, the selected MRs are also farther from the RAs, and hence suffering higher pathloss. Additionally, when each $\theta_{worst}$ direction of a cell has a cell-edge MS, each selected MR encounters an increased interference impact compared to the scenario of the $\theta_{best}$ direction. However, in terms of the effective throughput at the cell-edge, it is observed from Fig.~\ref{fig:throughput_improved_MR_selection_worst_direction} that apart from the high-complexity soft ML based MUD, all the non-cooperative MUDs effectively fail. This is because in the $\theta_{worst}$ direction, the BER performance of the non-cooperative MMSE-OSIC and of the non-cooperative PDA is so poor that hardly any packet can be successfully decoded. Additionally, in certain parts of the cell-edge area the SC-PDA based MR-FFR-DAS schemes indeed achieve a significantly higher effective throughput than the non-cooperative FFR-DAS that employs either the MMSE-OSIC or the PDA. However, their achievable effective throughput still remains much lower than that of the non-cooperative FFR-DAS employing the soft ML based MUD, especially when the cell-edge MSs approach the cell-edge boundary. 
 \begin{figure}[tbp]
     \centering
     {
         \includegraphics[width=\linewidth]{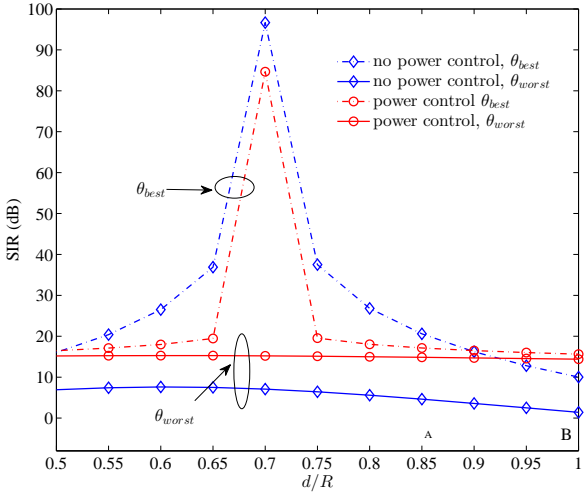}
     } 
     \caption{The average SIR experienced by cell-edge MSs of the MR-FFR-DAS scheme operating both with and without power control in the $\theta_{best}$ and $\theta_{worst}$ directions.}
     \label{fig:SIR_PC}
 \end{figure}

\subsubsection{Power Control in $\theta_{best}$ and $\theta_{worst}$ Directions}
\label{sec_power_control} 
 \begin{figure*}[t]
     \centering
     \subfigure[]{
      \label{fig:BER_power_control}
      \includegraphics[width=0.48\linewidth]{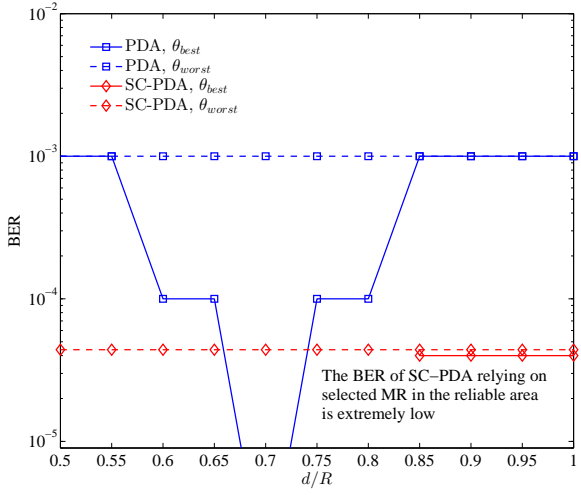}
     }
     \subfigure[]{
      \label{fig:throughput_power_control}
     \includegraphics[width=0.48\linewidth]{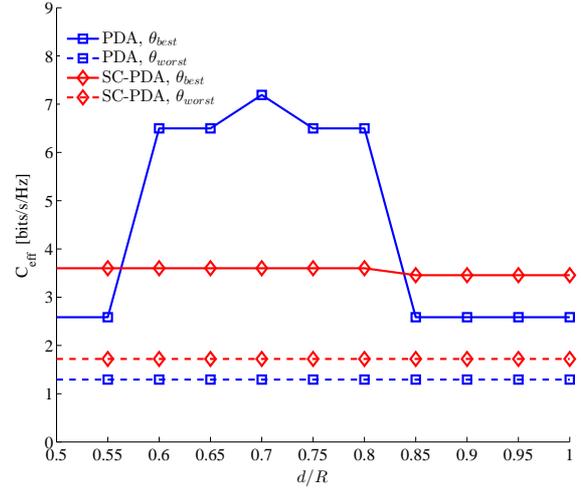}	  
     }
     \caption{BER and effective throughput of the non-cooperative FFR-DAS and the MR-FFR-DAS schemes operating with power control in the cell-edge area, when applying the non-cooperative PDA and the SC-PDA based MUD techniques, respectively. Both the $\theta_{best}$ and  $\theta_{worst}$ directions are evaluated.}
     \label{fig:PC_PDACPDA}
 \end{figure*}
We have observed from Fig.~\ref{fig:MR_selection_reliable_area_best_direction} and Fig.~\ref{fig:MR_selection_reliable_area_worst_direction} that although the BER of the MSs in the FFR-DAS based schemes has been substantially improved in most of the cell-edge area with the aid of the SC-PDA based MUD invoking MRs, the BER of MSs roaming very close to the cell-edge boundary (e.g. at the location where $d/R = 1$) remains higher than $10^{-3}$, even when activating an MR in the ``reliable area'' of the $\theta_{best}$ scenario. As a result, in terms of the effective throughput calculated relying on the packet error rate (PER) while assuming the packet length of 1024 bits, the cell-edge MSs are still poorly supported by the FFR-DAS based schemes. As a further remedy, we employ power control for improving the SIR for the cell-edge MSs roaming far away from the RAs.

The average SIR experienced by cell-edge MSs at the $\theta_{best}$ and $\theta_{worst}$ directions is recorded both in the absence and in the presence of power control, as  shown in Fig.~\ref{fig:SIR_PC}. For the scenario of the $\theta_{best}$ direction, we can see that when the MSs are roaming close to the RAs in the cell-edge area, i.e. $d/R \rightarrow 0.7$, a high average SIR of $\mathsf{SIR}_{MS}\geq 20$ dB is maintained without increasing the transmit power $P_t$. By contrast, when power control is used, a similarly high  average SIR of $\mathsf{SIR}_{MS}^{'}\geq 20$ dB is also experienced by the MSs roaming close to the RAs. However, in this context the average SIR value obtained is in general slightly lower than that recorded in the absence of power control. This is because in the system where multiple users share the same frequency bandwidth simultaneously, power control is an inefficient technique from a sum-capacity perspective. To elaborate a little further, the interference imposed by the other co-channel MSs on the MS of interest may be increased due to increasing their respective transmit power for the sake of maintaining their link-quality. However, in the absence of power control, the SIR of the MSs roaming far away from the RAs is seriously reduced in the cell-edge area, especially in the $\theta_{worst}$ direction, where we have $0\text{dB} < \mathsf{SIR}_{MS} < 10\text{dB}$. As an improvement of user-fairness, when applying power control in the $\theta_{worst}$ direction, the cell-edge MSs that suffer from a low SIR of $\mathsf{SIR}_{MS}< 10\text{dB}$ have seen their average SIR increased to $\mathsf{SIR}_{MS}^{'} \approx 15 \text{dB}$, as evidenced by Fig.~\ref{fig:SIR_PC}. 
\begin{figure}[tbp]
\centering
\includegraphics[width=\linewidth]{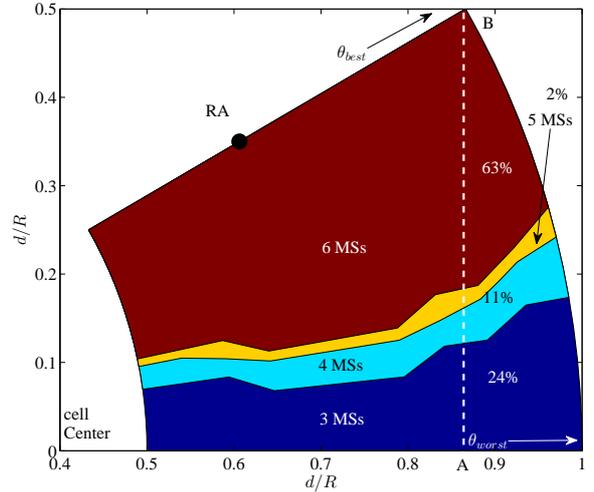} %
\caption{The number of simultaneously supported MSs satisfying the QoS of $\mathsf{SIR} > 15$ dB.}
\label{fig:QoS}
\end{figure} 

Both the BER and the effective throughput of the MSs supported by the non-cooperative PDA based FFR-DAS and the SC-PDA based MR-FFR-DAS  along the $\theta_{best}$ and $\theta_{worst}$ directions in the presence of power control are characterized in Fig.~\ref{fig:PC_PDACPDA}. More specifically, when using power control in the absence of MRs, namely when using the non-cooperative PDA, the MSs roaming in the cell-edge area are capable of achieving an improved BER of about $10^{-3}$ in the $\theta_{worst}$ direction. By contrast, when invoking power control and selecting the MRs from the ``reliable area'' for the SC-PDA, the MSs roaming in the cell-edge area are capable of achieving $\mathsf{BER} < 10^{-4}$ even when they are in the $\theta_{worst}$ direction and quite close to the cell-edge boundary, as shown in Fig.~\ref{fig:BER_power_control}. Additionally, when considering the effective throughput, we observe from Fig.~\ref{fig:throughput_power_control} that the SC-PDA based MR-FFR-DAS significantly outperforms the PDA based non-cooperative FFR-DAS in most of the cell-edge area, including the cell-edge boundary of both the $\theta_{best}$ and $\theta_{worst}$ directions. Meanwhile, in the SC-PDA based MR-FFR-DAS, all the cell-edge MSs have achieved a similarly high effective throughput, which implies having an appealing user-fairness.  
 \begin{figure*}[tbp]
     \centering
     \subfigure[]{
      \label{fig:CAS_v_DAS_BER}
      \includegraphics[width=0.48\linewidth]{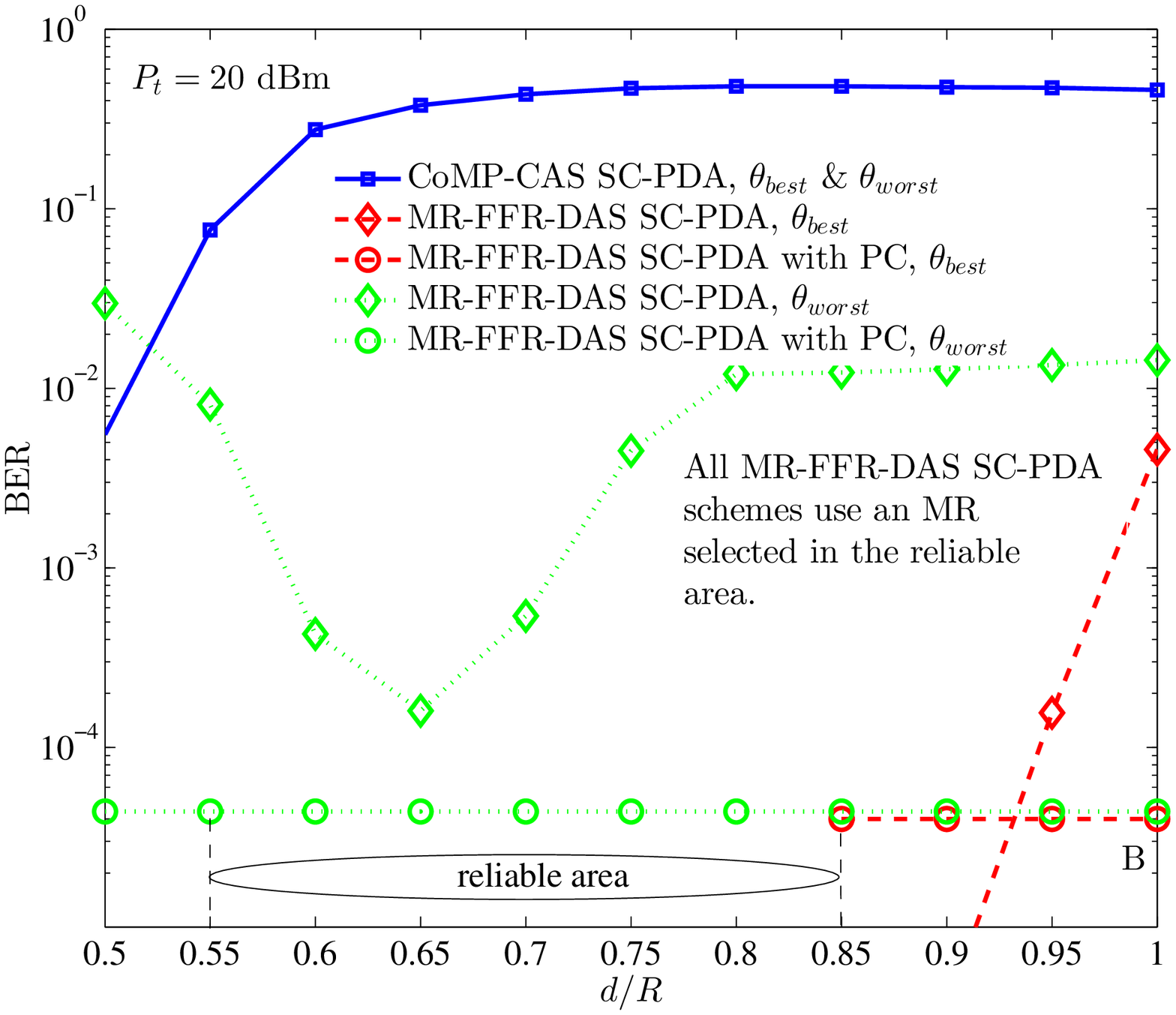}
     }
     \subfigure[]{
      \label{fig:CAS_DAS_throuput}
     \includegraphics[width=0.48\linewidth]{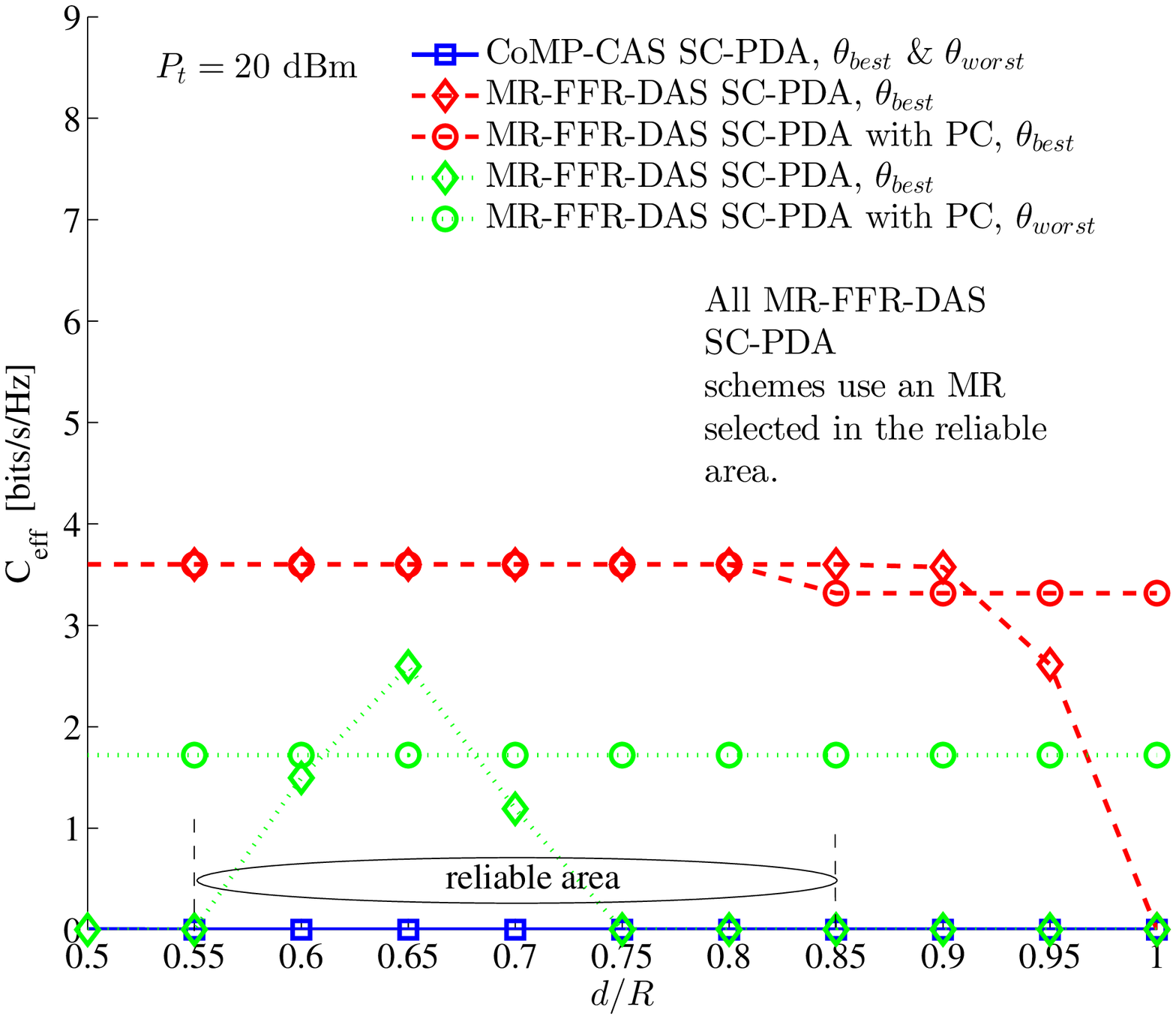}	  
     }
     \caption{The SC-PDA aided CoMP-CAS versus the SC-PDA aided MR-FFR-DAS in terms of their (a) BER and (b) effective throughput in the cell-edge area.}
     \label{fig:CAS_v_DAS}
 \end{figure*} 
 
In order to more clearly demonstrate the benefits of jointly using the power control and the MR-aided SC-PDA detector, the QoS distribution across the entire cell-edge area in shown Fig.~\ref{fig:QoS}, where we define the minimum QoS required as $\mathsf{SIR}_{MS}^{'}> 15\text{dB}$. Our observation area is defined as the $30^{\circ}$ sector spanning the $\theta_{best}$ and $\theta_{worst}$ directions, as visualized in Fig.~\ref{fig:Tpgy_DAS}, where a total of $N_m =6$ MSs are simultaneously supported across the entire cell-edge area. We can see that when the MSs roam close to the RAs, all the $6$ MSs are capable of achieving the QoS target of $\mathsf{SIR}_{MS}^{'}> 15\text{dB}$ with the aid of power control, which corresponds to $63\%$ of the entire cell-edge area, as shown in Fig.~\ref{fig:QoS}. When the MSs are roaming close to the $\theta_{worst}$ direction, the number of MSs achieving the QoS target of $\mathsf{SIR}_{MS}^{'}> 15\text{dB}$ is reduced to $3$, which corresponds to $24\%$ of the 
entire cell-edge area.

Finally, in Fig. \ref{fig:CAS_v_DAS_BER} and Fig.~\ref{fig:CAS_DAS_throuput} we compared the SC-PDA aided CoMP-CAS and the SC-PDA aided MR-FFR-DAS in terms of their achievable cell-edge BER and cell-edge effective throughput, respectively. These results have clearly shown that the SC-PDA aided MR-FFR-DAS constitutes a more promising solution for improving the cell-edge performance of interference-limited multi-cell systems.      

\section{Conclusions}
\label{secUL_cn}
In this paper, we studied the achievable uplink cell-edge performance of four multi-cell system architectures, including the non-cooperative CAS, the BS-cooperation aided CoMP-CAS, the non-cooperative FFR-DAS and the MR-FFR-DAS architectures. By benchmarking against three representative non-cooperative MUD schemes, we demonstrated that typically both the CoMP-CAS and the MR-FFR-DAS relying on the proposed SC-PDA receiver significantly outperform their respective non-cooperative counterparts in terms of the BER of the cell-edge MSs. We revealed, however that the performance gain achieved by the CoMP-CAS may be eroded when the transmit power of the cell-edge MSs is low, and the MR-FFR-DAS is in general more effective in supporting cell-edge MSs than the CoMP-CAS. We also demonstrated that the cell-edge BER of the MR-FFR-DAS may be further reduced by using judicious MR selection, especially for the MSs roaming in the "worst-case direction" or close to the cell-edge boundary. On the other hand, when considering the cell-edge effective throughput calculated relying on the packet error rate, we showed that the SC-PDA aided MR-FFR-DAS architecture does not always outperform its non-cooperative counterpart, since the former invokes two time-slots for completing a single MS-transmission. Furthermore, we demonstrated that the effective throughput at the cell-edge and the fairness amongst the cell-edge MSs in the SC-PDA aided MR-FFR-DAS may be significantly improved by using multi-user power control. As a result, for low/moderate MS transmit powers, the proposed SC-PDA aided MR-FFR-DAS scheme is capable of achieving both a significantly better BER and an improved effective throughput across the entire cell-edge area than the CoMP-CAS scheme.

\bibliography{xxy} 

\begin{IEEEbiography}[{\includegraphics[width=1in,height=1.25in,clip,keepaspectratio]{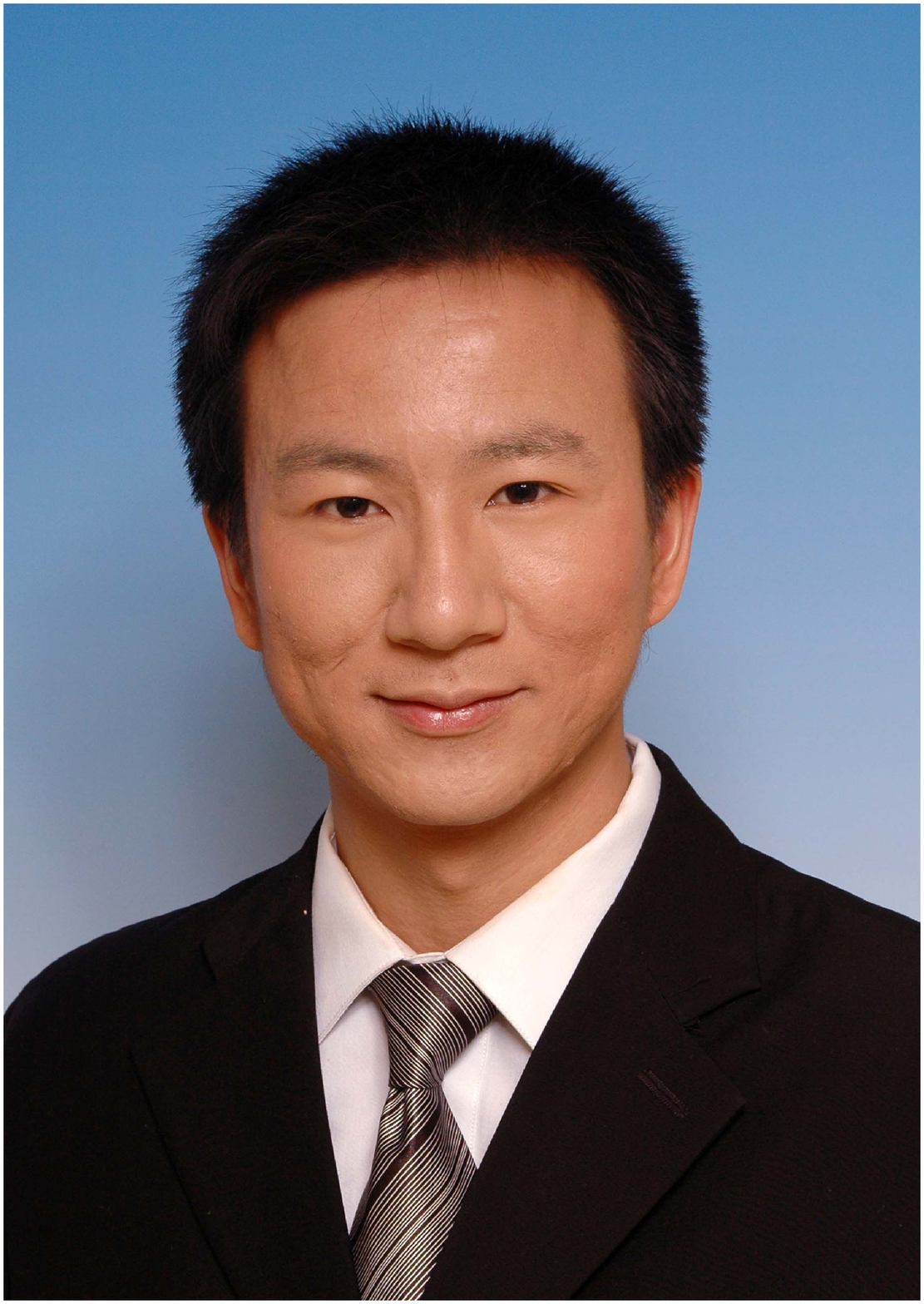}}]{Shaoshi Yang}
(S'09-M'13)  received
the B.Eng. Degree in Information Engineering
from Beijing University of Posts and
Telecommunications (BUPT), China, in 2006, the first Ph.D. Degree in Electronics and Electrical Engineering from University of Southampton, U.K., in 2013, and a second Ph.D. Degree in Signal and Information Processing from BUPT in 2014. Since 2013 he has been a Postdoctoral Research Fellow in
University of Southampton, U.K, and from 2008 to 2009, he was
an Intern Research Fellow with the Intel Labs China, Beijing, where he focused on
Channel Quality Indicator Channel design for mobile WiMAX (802.16m).
His research interests include MIMO signal processing, green radio, heterogeneous networks, cross-layer interference management, convex optimization and its applications. He has published
in excess of 30 research papers on IEEE journals and conferences.

Shaoshi has received a number of academic and research awards, including the PMC-Sierra Telecommunications Technology Scholarship at BUPT, the Electronics and Computer Science (ECS) Scholarship of University of Southampton and the Best PhD Thesis Award of BUPT. He serves as a TPC member of a number of IEEE conferences and journals, including IEEE ICC, PIMRC, ICCVE, HPCC and IEEE Journal on Selected Areas in Communications. He is also a Junior Member of the Isaac Newton Institute for Mathematical Sciences, Cambridge University, UK. (https://
sites.google.com/site/shaoshiyang/)
\end{IEEEbiography}

\begin{IEEEbiography}[{\includegraphics[width=1in,height=1.25in,clip,keepaspectratio]{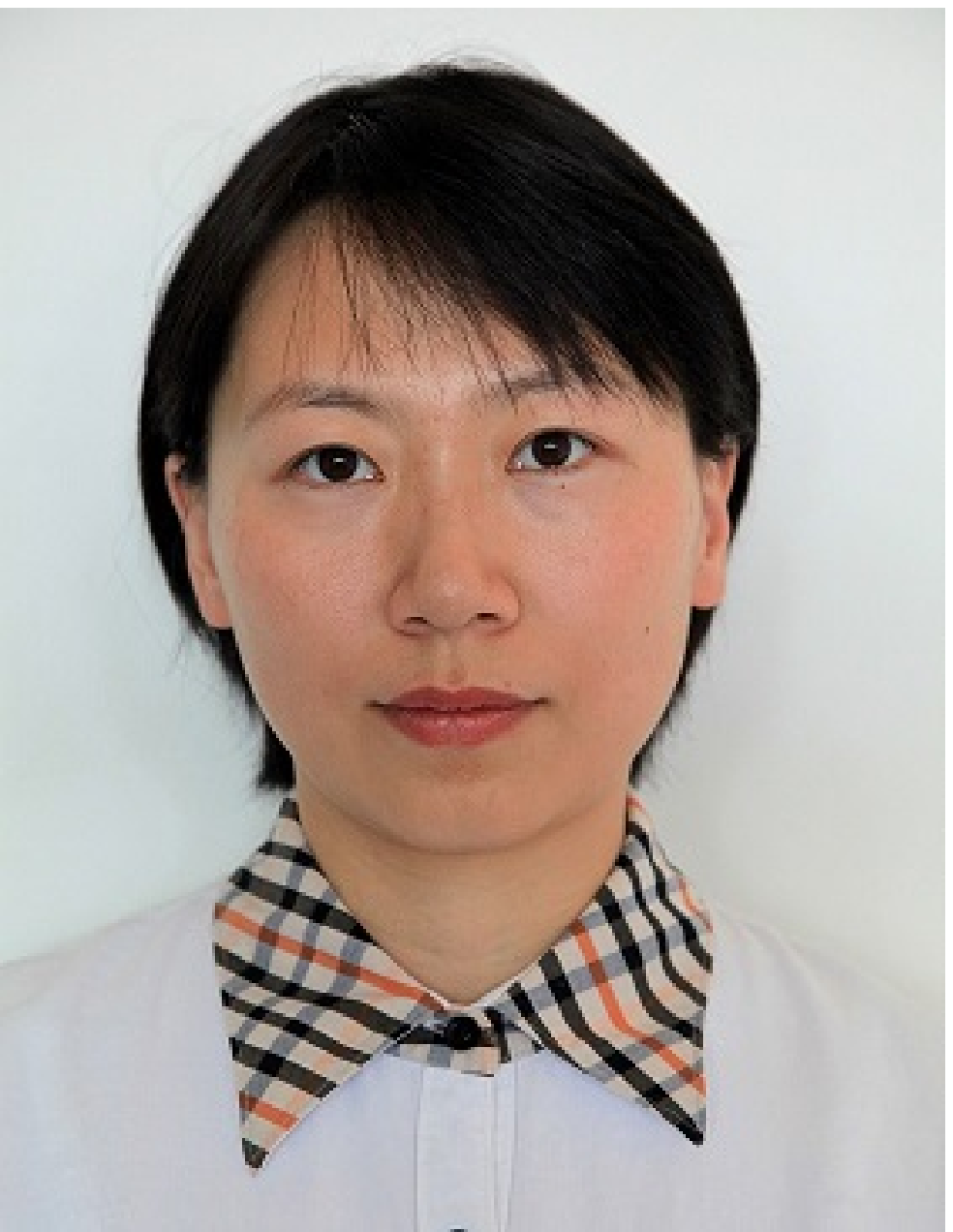}}]{Xinyi Xu} received her B.Eng. degree in Electronic Information and Ph.D. degree in communications \& Information System from Wuhan University, China, in 2004 and 2014, respectively. She also received her Ph.D. degree in wireless communications from the University of Southampton, United Kingdom, in 2013, which was sponsored by the UK/China scholarships for excellence programme since 2008. She is now working in the China Academy of Electronics and Information Technology. Her research interests include future network and Integrated space-around Network.
\end{IEEEbiography}

\begin{IEEEbiography}[{\includegraphics[width=1in,height=1.25in,clip,keepaspectratio]{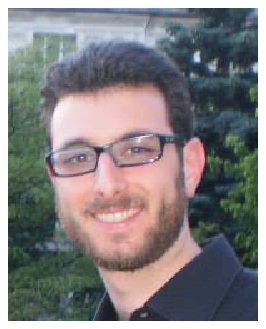}}]{Dimitrios Alanis}
(S'13) received the M.Eng. degree in Electrical and Computer Engineering from the Aristotle University of Thessaloniki  in 2011 and the M.Sc. degree in Wireless Communications from the University of Southampton in 2012. He is currently working towards the PhD degree with the Southampton Wireless Group, School of Electronics and Computer Science of the University of Southampton. 
His research interests include quantum computation and quantum information theory, quantum search algorithms, cooperative communications, resource allocation for self-organizing networks, bio-inspired optimization algorithms and classical and quantum game theory.
\end{IEEEbiography}

\begin{IEEEbiography}[{\includegraphics[width=1in,height=1.25in,clip,keepaspectratio]{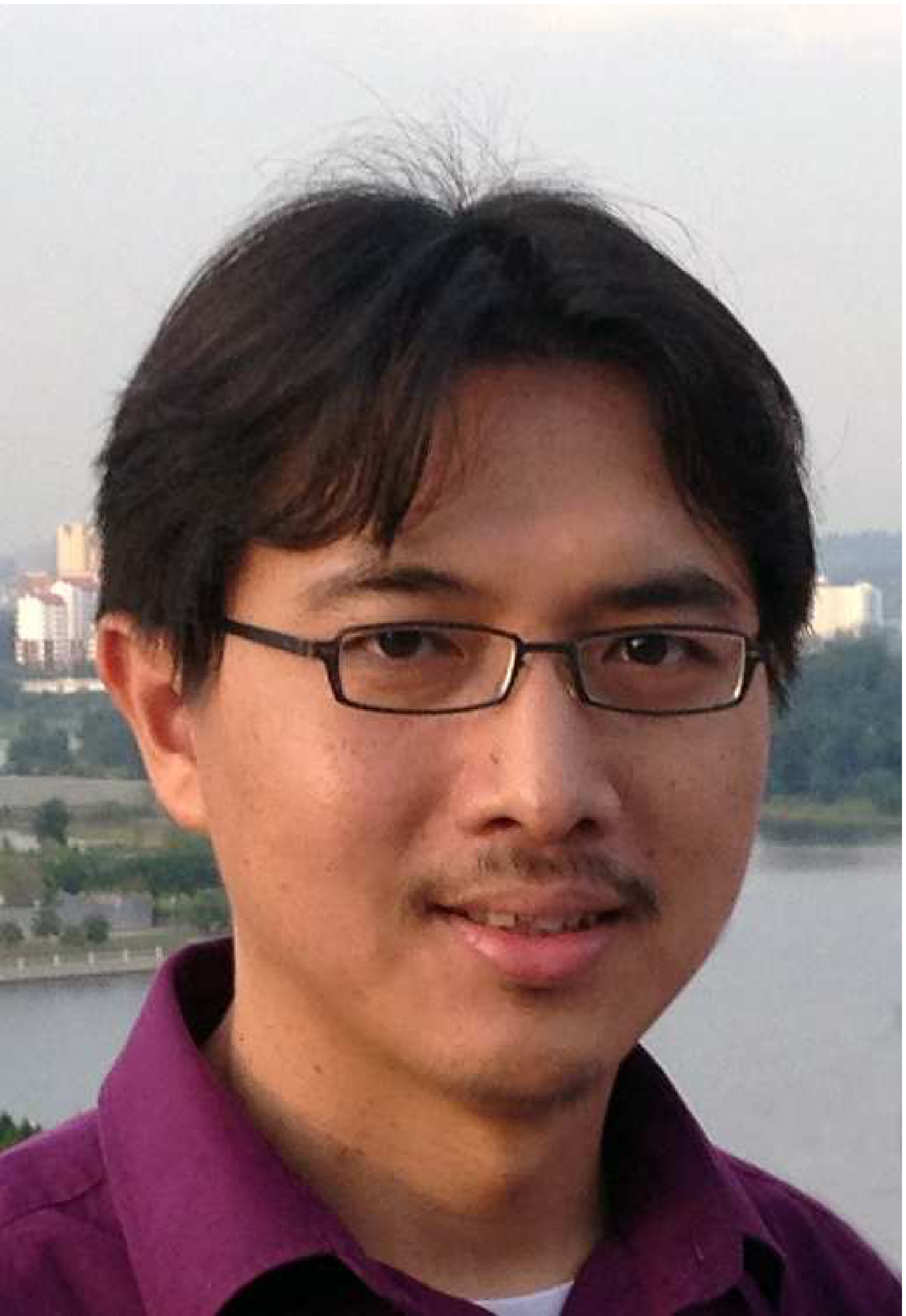}}]{Soon Xin Ng} (S'99-M'03-SM'08) received the B.Eng. degree (First class)
in electronic engineering and the Ph.D. degree in telecommunications
from the University of Southampton, Southampton, U.K., in 1999 and 2002,
respectively. From 2003 to 2006, he was a postdoctoral research fellow
working on collaborative European research projects known as SCOUT,
NEWCOM and PHOENIX. Since August 2006, he has been a member of academic
staff in the School of Electronics and Computer Science, University of
Southampton. He is involved in the OPTIMIX and CONCERTO European
projects as well as the IU-ATC and UC4G projects. He is currently an
Associate Professor in telecommunications at the University of Southampton.

His research interests include adaptive coded modulation, coded
modulation, channel coding, space-time coding, joint source and channel
coding, iterative detection, OFDM, MIMO, cooperative communications,
distributed coding, quantum error correction codes and joint
wireless-and-optical-fibre communications. He has published over 180
papers and co-authored two John Wiley/IEEE Press books in this field. He
is a Senior Member of the IEEE, a Chartered Engineer and a Fellow of the
Higher Education Academy in the UK.
\end{IEEEbiography}

\begin{IEEEbiography}[{\includegraphics[width=1in,height=1.25in,clip,keepaspectratio]{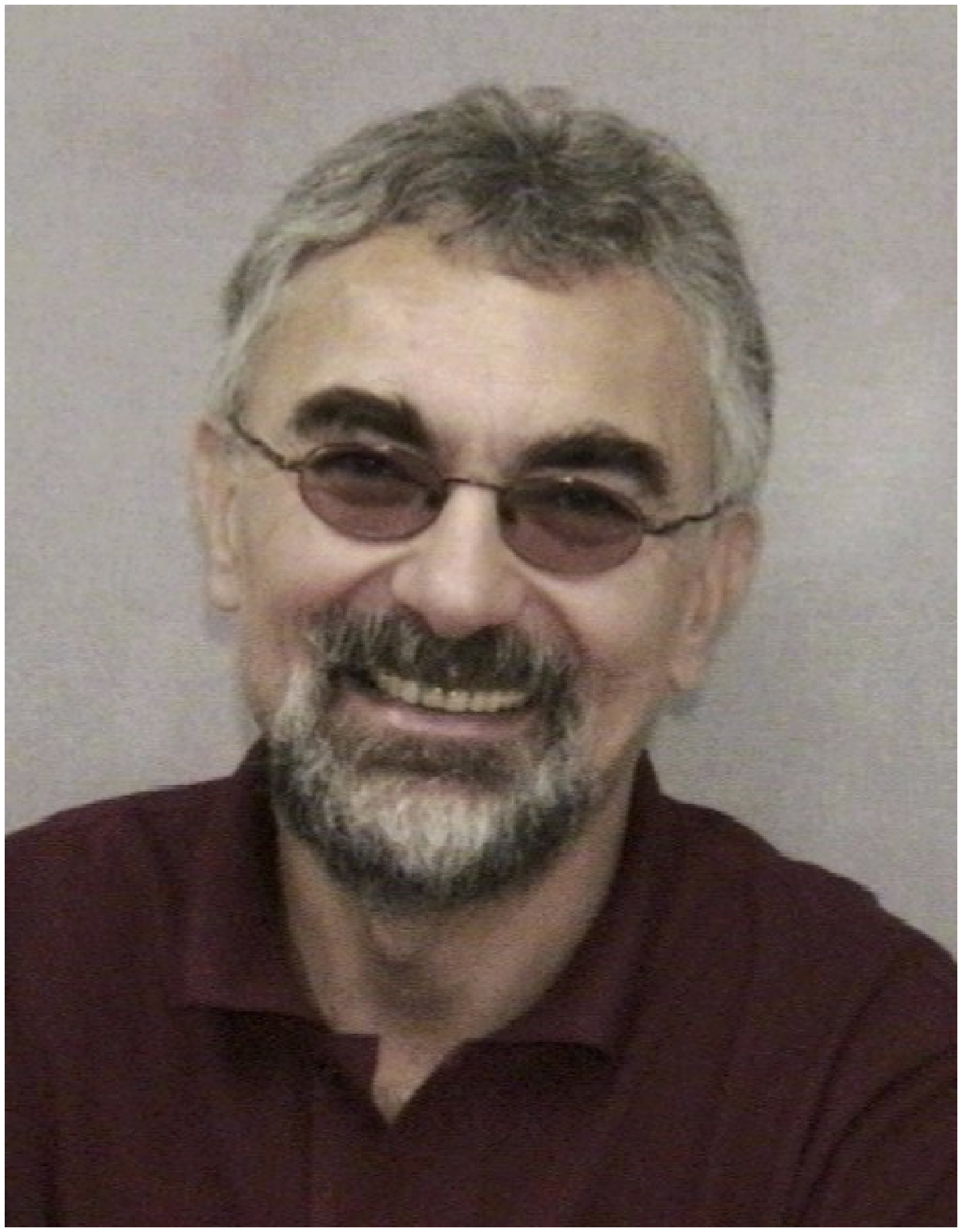}}]{Lajos Hanzo}
(M'91-SM'92-F'04) received his degree in electronics in
1976 and his doctorate in 1983.  In 2009 he was awarded the honorary
doctorate ``Doctor Honoris Causa'' by the Technical University of
Budapest.  During his 38-year career in telecommunications he has held
various research and academic posts in Hungary, Germany and the
UK. Since 1986 he has been with the School of Electronics and Computer
Science, University of Southampton, UK, where he holds the chair in
telecommunications.  He has successfully supervised about 100 PhD students,
co-authored 20 John Wiley/IEEE Press books on mobile radio
communications totalling in excess of 10 000 pages, published 1400+
research entries at IEEE Xplore, acted both as TPC and General Chair
of IEEE conferences, presented keynote lectures and has been awarded a
number of distinctions. Currently he is directing a 100-strong
academic research team, working on a range of research projects in the
field of wireless multimedia communications sponsored by industry, the
Engineering and Physical Sciences Research Council (EPSRC) UK, the
European Research Council's Advanced Fellow Grant and the Royal
Society's Wolfson Research Merit Award.  He is an enthusiastic
supporter of industrial and academic liaison and he offers a range of
industrial courses. 

Lajos is a Fellow of the Royal Academy of Engineering, of the Institution
of Engineering and Technology, and of the European Association for Signal
Processing. He is also a Governor of the IEEE VTS.  During
2008 - 2012 he was the Editor-in-Chief of the IEEE Press and a Chaired
Professor also at Tsinghua University, Beijing. He 
has 20 000+ citations. For further information on research in progress and associated
publications please refer to http://www-mobile.ecs.soton.ac.uk 
\end{IEEEbiography}

\newpage

\end{document}